\documentclass[final,5p,times,twocolumn]{elsarticle}

\usepackage{amssymb}
\usepackage{amsmath}

\usepackage[table]{xcolor}
\colorlet{shadecolor}{gray!15}

\usepackage{etoolbox}
\robustify\bfseries
\usepackage{graphicx}
\usepackage{amsmath,amssymb,graphicx}
\usepackage[]{algorithm2e}
\usepackage{tabularx}
\usepackage{wasysym}
\usepackage{color,soul}
\usepackage{balance}
\usepackage{breqn}
\usepackage{makecell}
\usepackage{booktabs}
\usepackage{multirow}
\usepackage{siunitx}
\usepackage{xcolor}
\usepackage{hyperref}
\def\H{{\mathsf H}}
\def\T{{\mathsf T}}
\def\CC{{\mathbb C}}
\def\RR{{\mathbb R}}
\usepackage{caption}
\usepackage{arydshln}
\usepackage{enumitem}
\usepackage{mathtools}

\usepackage{lscape}
\usepackage{pdflscape}

\makeatletter
\newcommand*{\bigs}[1]{{\hbox{$\left#1\vbox to9\p@{}\right.\n@space$}}}
\makeatother

\usepackage{amssymb} %
\usepackage{pifont} %
\newcommand{\cmark}{\ding{51}}%
\newcommand{\xmark}{\ding{55}}%

\usepackage{tikz}

\newcommand{\ZQHL}[1]{#1} %
\newcommand{\ZQHLRtwo}[1]{{#1}} %

\journal{Neural Networks}

\usepackage{lineno}

\begin{document}

\begin{frontmatter}

\title{SuperM2M: Supervised and Mixture-to-Mixture Co-Learning for Speech Enhancement and \ZQHL{Noise-}Robust ASR}

\author[label1]{Zhong-Qiu Wang} %
\ead{wang.zhongqiu41@gmail.com}

\affiliation[label1]{organization={Department of Computer Science and Engineering}, addressline={Southern University of Science and Technology}, city={Shenzhen}, postcode={518055}, state={Guangdong}, country={P.R. China}}

\begin{abstract}
The current dominant approach for neural speech enhancement is based on supervised learning by using simulated training data.
The trained models, however, often exhibit limited generalizability to real-recorded data.
To address this, this paper investigates training enhancement models directly on real target-domain data.
We propose to adapt mixture-to-mixture (M2M) training, originally designed for speaker separation, for speech enhancement, by modeling multi-source noise signals as a single, combined source.
In addition, we propose a co-learning algorithm that improves M2M with the help of supervised algorithms.
When paired close-talk and far-field mixtures are available for training, M2M realizes speech enhancement by training a deep neural network (DNN) to produce speech and noise estimates in a way such that they can be linearly filtered to reconstruct the close-talk and far-field mixtures.
This way, the DNN can be trained directly on real mixtures, and can leverage close-talk and far-field mixtures as a weak supervision to enhance far-field mixtures.
To improve M2M, we combine it with supervised approaches to co-train the DNN, where mini-batches of real close-talk and far-field mixture pairs and mini-batches of simulated mixture and clean speech pairs are alternately fed to the DNN, and the loss functions are respectively (a) the mixture reconstruction loss on the real close-talk and far-field mixtures and (b) the regular enhancement loss on the simulated clean speech and noise.
We find that, this way, the DNN can learn from real and simulated data to achieve better generalization to real data.
We name this algorithm SuperM2M (supervised and mixture-to-mixture co-learning).
Evaluation results on the CHiME-4 dataset show its effectiveness and potential.
\end{abstract}


\end{frontmatter}

\section{Introduction}

Deep learning has dramatically advanced speech enhancement \cite{WDLreview}.
The current dominant approach is based on supervised learning, where clean speech is synthetically mixed with noises in simulated reverberant conditions to create paired clean speech and noisy-reverberant mixtures for training neural speech enhancement models in a supervised, discriminative way to predict the clean speech from its paired mixture \cite{WDLreview}.
Although showing strong performance in matched simulated test conditions \cite{Chen2016c, Ephrat2018, Luo2019ConvTasNet, Luo2020DPRNN, Zmolikova2023SPM, Kavalerov2019, Nachmani2020, Zeghidour2020, Chen2020CSS, Xu2020SpEx, Wang2020CSMDereverbJournal, Wang2020chime, Wang2020css, Tan2022NSF, Tesch2021, Zhang2021ADLMVDR, Wang2022GridNetjournal, Chetupalli2023EENDEDASeparation, Zheng2023sixty, Saijo2023, Pons2024GASS, Quan2024SpatialNet, Zhang2023USES, Zhang2024USES2}, the trained models often exhibit limited generalizability to real-recorded data \cite{WDLreview, Pandey2020CrossCorpus, Zhang2021ClosingGap, Tzinis2022REMIXT, Tzinis2022AudioScopeV2, Cox2023ClaritySP2023, Leglaive2023CHiME7UDASE, Cornell2023, HaebUmbach2019SPM, Zhang2023USES, Zhang2024USES2, Yang2024}, largely due to mismatches between simulated training and real-recorded test conditions.

A possible way to improve the generalizability, we think, is to have the model see, and learn to model, real-recorded target-domain mixtures during training.
This, however, cannot be applied in a straightforward way, since the clean speech at each sample of the real mixtures cannot be annotated or computed in an easy way.
As a result, there lacks a good sample-level supervision for real mixtures, unlike simulated mixtures where a sample-level supervision is readily available.

During data collection, multiple far-field microphones are usually utilized to record target speakers.
In UNSSOR \cite{Wang2023UNSSOR}, our recent algorithm proposed for unsupervised speaker separation, we find that the mixture signal recorded by each far-field microphone can be leveraged as a \textit{weak supervision} for training DNNs to separate speakers.
The idea is that each far-field mixture can be utilized as a constraint to regularize speaker estimates.
That is, the speaker estimates, produced by using the mixtures captured at a subset of microphones as input, should be capable of being leveraged to reconstruct the mixture captured by each microphone.

On the other hand, during data collection, besides using far-field microphones to record target speech, a close-talk microphone is often placed near the target speaker to collect its close-talk speech (e.g., in the CHiME \cite{Barker2018CHiME5}, AMI \cite{McCowan2006}, AliMeeting \cite{Yu2022M2MeT}, and MISP \cite{Wu2024MISP2024} setup).\footnote{Close-talk speech is almost always recorded together with far-field speech in speech separation and recognition datasets, as it is much easier for humans to annotate word transcriptions and speaker activities based on close-talk recordings (where the target speech is very strong) than far-field recordings.}
Although the close-talk microphone can also pick up non-target signals, the recorded close-talk mixture typically has a much higher signal-to-noise ratio (SNR) of the target speaker than any far-field mixtures.
In our recent mixture to mixture (M2M) algorithm \cite{Wang2024M2MSPL}, which builds upon UNSSOR, we find that, besides far-field mixtures, the close-talk mixture can also be leveraged as a \textit{weak supervision} for training DNNs to separate mixed speakers.

To leverage the weak-supervision in far-field and close-talk mixtures for separation, two difficulties need to be solved.
First, far-field mixtures contain multiple sources and are not clean, and due to the contamination of the other sources \cite{McCowan2006, Barker2018CHiME5, Watanabe2020CHiME6}, close-talk mixtures are often not clean enough.
Second, each speaker's image in the close-talk mixture is not time-aligned with its image in each far-field mixture, and each speaker's images in different far-field mixtures are also not time-aligned with each other.
As a result, close-talk and far-field mixtures cannot be naively used as the training targets for training speaker separation models.
To overcome the two difficulties, we have recently proposed UNSSOR in a conference publication \cite{Wang2023UNSSOR} and M2M in a letter submission \cite{Wang2024M2MSPL}.
The idea is that, at each training step, we can (a) feed a far-field mixture to a DNN to produce an estimate for each speaker; and (b) regularize the speaker estimates such that they can be linearly filtered via multi-frame linear filtering to reconstruct the close-talk and far-field mixtures.
This way, the first difficulty is addressed by having the filtered speaker estimates to respectively approximate (i.e., explain) the speaker images in each mixture, and the second difficulty is addressed by multi-frame linear filtering.

Although UNSSOR and M2M are capable of being trained directly on real-recorded mixtures (i.e., not requiring the availability of clean speech), they have been only trained and evaluated on simulated mixtures \cite{Wang2023UNSSOR, Wang2024M2MSPL}.
It is yet unknown (a) whether they are effective on real data; and (b) whether they can lead to better generalization to real data, compared with the current dominant purely-supervised approaches, which train models only on simulated data.
We emphasize that these two concerns are very reasonable, since, on real data, the physical models hypothesized in UNSSOR and M2M are expected to be much less satisfied.
For example, there could be microphone synchronization errors, microphone failures, different frequency responses in different microphones, signal clipping, slight speaker and array movement, non-linear filter relationships among speaker images at different microphones, distortions to target speech caused by real microphones, etc.
These issues can potentially pose difficulties for UNSSOR and M2M.
In addition, UNSSOR and M2M were designed for separating mixed speakers.
It is unclear whether they would be effective for single-speaker speech enhancement and robust automatic speech recognition (ASR), where suppressing non-target signals (such as noises) is a major concern.
Furthermore, UNSSOR and M2M assume stationary, weak Gaussian noises in their physical models.
It is unclear whether they can deal with strong, non-stationary noises, which could contain an unknown number of diffuse and directional sources.

In this context, we propose to extend UNSSOR \cite{Wang2023UNSSOR} and M2M \cite{Wang2024M2MSPL} for speech enhancement and robust ASR, where we train and evaluate the proposed algorithms not only on simulated data but also on real data.\footnote{\ZQHL{This long-form paper is a journal expansion of our preliminary work in UNSSOR \cite{Wang2023UNSSOR} and M2M \cite{Wang2024M2MSPL}.}}
Our major goal is to show whether the resulting algorithms can yield better generalization to real data than purely-supervised models trained on simulated data, a demonstration that is missing in UNSSOR \cite{Wang2023UNSSOR} and M2M \cite{Wang2024M2MSPL}.
Without this demonstration, the evidence supporting whether this un- and weakly-supervised line of research is worth investigating would be lacking, especially considering the dominance and simplicity of purely-supervised approaches based on simulated training data.
We summarize the key contributions of this paper as follows:
\begin{itemize}[leftmargin=*,noitemsep,topsep=0pt]
\item We are the first proposing to leverage close-talk and far-field mixtures as weak supervision for speech enhancement, a task different from speaker separation.
\item Considering noise sources as a single, combined source, we formulate the training of speech enhancement models on real data as solving a blind deconvolution problem, following the formulations in UNSSOR \cite{Wang2023UNSSOR} and M2M \cite{Wang2024M2MSPL} designed for speaker separation.
\item \ZQHL{To account for the case when close-talk mixture is not time-synchronized with its paired far-field mixtures, we propose an on-the-fly filter tap estimation algorithm that can deal with the time-synchronization issue.}
\item We propose SuperM2M, a co-learning strategy which trains the same DNN model by alternating between M2M training on real data and supervised learning on simulated data.
This way, M2M can benefit from massive simulated training data, especially when the real training data is scarce.
In addition, we find that this strategy can help mitigate the weaknesses of UNSSOR and M2M on \ZQHL{source permutation, frequency permutation, and source ambiguity (all of which will be detailed in Section \ref{m2m_weakness})}.
\end{itemize}
We validate SuperM2M on the CHiME-4 dataset \cite{Vincent2016CHiME4Analysis}, which is consisted of simulated and challenging real-recorded mixtures and is currently the major benchmark for evaluating robust ASR and speech enhancement algorithms.
In our experiments, state-of-the-art ASR and enhancement performance is obtained.
The evaluation results suggest that:
\begin{itemize}[leftmargin=*,noitemsep,topsep=0pt]
\item SuperM2M can effectively learn from real mixtures and leverage the weak supervision afforded by real close-talk and far-field mixtures.
\item The co-learning strategy can significantly improve the generalizability of purely-supervised models trained on simulated data to real data.
\end{itemize}
The evaluation results provide an evidence supporting the strong potential of our un- and weakly-supervised line of research for speech enhancement. 
A sound demo is provided in the link below.\footnote{\scriptsize \url{https://zqwang7.github.io/demos/SuperM2M_demo/index.html}}

\section{Related Work}\label{related_work}

SuperM2M is related to other work in five major aspects.

\subsection{Frontend Enhancement and Robust ASR}

Leveraging neural speech enhancement as a frontend processing to improve the robustness of backend ASR systems to noise, reverberation and competing speech has been a long-lasting research topic \cite{HaebUmbach2019SPM, Haeb-Umbach2020}.
Although dramatic progress has been made in neural speech enhancement \cite{WDLreview, Zheng2023sixty}, directly feeding the immediate estimate produced by DNN-based enhancement models for ASR has had limited success, largely for two reasons: (a) enhancement DNNs, which can suppress non-target signals aggressively, often incur speech distortion detrimental to ASR; and (b) enhancement DNNs are often trained on simulated data, which inevitably mismatches real data, and this mismatch further aggravates the speech distortion problem.
Through years of efforts, robust ASR approaches have gradually converged to (a) leveraging DNN estimates to derive linear beamforming results for ASR \cite{Heymann2015, Zhang2017a, Boeddecker2018GSS}; and (b) jointly training ASR models with enhancement models \cite{Narayanan2015JointTraining, Wang2016JointTrainingASR, Heymann2017BeamNet, Chang2019MIMOSpeech}.
These two approaches aim at improving robust ASR performance.
Their enhancement modules usually do not produce sufficiently accurate estimation of target speech.
For example, linear beamforming is known to introduces little speech distortion but it has limited capabilities at suppressing non-target signals (especially when the number of microphones is limited and when the non-target signals are diffuse) \cite{Gannot2017}.
Another example is that jointly training enhancement models with ASR models often degrades the performance of the enhancement models on realistic mixtures \cite{Masuyama2023SS}.

Differently, in this paper we aim at building neural speech enhancement models whose immediate estimate \underline{\textit{itself}} can have low distortion to target speech and high reduction to non-target signals, especially on real test data.
We find that, on the challenging real test data of CHiME-4 \cite{Vincent2016CHiME4Analysis}, the immediate output of SuperM2M bears low distortion to target speech and high reduction to non-target signals, and feeding it directly to strong ASR models for recognition yields strong performance.

\subsection{Generalizability of Supervised Models to Real Data}

Improving the generalizability of neural speech enhancement models to real data has received decade-long efforts.
The current dominant approach \cite{Chen2016c, WDLreview, Zhang2023USES, Zhang2024USES2} is to train supervised models on large-scale synthetic data, which is simulated in a way to cover as many variations (that could happen in real test data) as possible.
However, the success has been limited, largely due to the current simulation techniques being not good enough at generating simulated mixtures as realistic as real mixtures.
This can be observed from recent speech enhancement and ASR challenges.
In the Clarity enhancement challenge \cite{Cox2023ClaritySP2023}, all the teams scored well on simulated data failed on real data.
In CHiME-3/4 \cite{Vincent2016CHiME4Analysis}, in the multi-channel cases, all the top teams use conventional beamformers (although with signal statistics estimated based on DNN estimates) as the only frontend, and in the single-channel cases, frontend enhancement often degrades ASR performance compared to not using any enhancement (assuming no joint frontend-backend training) \cite{Chang2022E2EIntegration}.
In CHiME-\{5,6,7\} \cite{Watanabe2020CHiME6} and M2MeT \cite{Yu2022M2MeT}, almost all the teams adopt guided source separation \cite{Boeddecker2018GSS}, a signal processing algorithm, as the only frontend.

Since the current simulation techniques are not satisfactory enough, a possible way to improve the generalizability to real data, we think, is to train enhancement models directly on real data.

\subsection{Unsupervised Speech Separation}

To model real data, unsupervised neural speech separation algorithms (such as MixIT \cite{Wisdom2020MixIT}, ReMixIT \cite{Tzinis2022REMIXT}, NyTT \cite{Fujimura2021}, Neural FCA \cite{Bando2021NeuralFCA}, RAS \cite{Aralikatti2022RAS}, UNSSOR \cite{Wang2023UNSSOR} and USDnet \cite{Wang2024USDnet}), which can train separation models directly on mixtures or synthetic mixtures of mixtures, have been proposed.
Due to their unsupervised nature, their performance could be limited due to not leveraging any supervision.
Meanwhile, many algorithms in this stream are only evaluated on simulated data and their effectiveness on real data and for robust ASR is unclear.
In contrast, we will show that SuperM2M works well on the challenging real data of CHiME-4.

\subsection{Semi-Supervised Speech Separation}

A promising direction, suggested by \cite{Sivaraman2022Adapttorealmeeting, Han2024Adapt} (and subsequent studies \cite{ZhangJISI2022MixIT, Hao2023}), is to combine supervised learning on simulated data and unsupervised learning on real data for model training, forming a semi-supervised approach.
The rationale is that supervised learning on massive simulated data offers an easy and feasible way for the model to learn to model speech patterns, and unsupervised learning on real data can help the model learn from real data.

SuperM2M follows this direction, but differs from \cite{Sivaraman2022Adapttorealmeeting, Han2024Adapt} in two major aspects.
First, SuperM2M leverages M2M \cite{Wang2024M2MSPL}, which builds upon UNSSOR \cite{Wang2023UNSSOR}, to model real data, while \cite{Sivaraman2022Adapttorealmeeting, Han2024Adapt} uses MixIT \cite{Wisdom2020MixIT}.
As is suggested in \cite{Wang2023UNSSOR, Wang2024USDnet}, UNSSOR based methods (a) avoid tricky (and often unrealistic) synthesis of mixtures of mixtures, which, on the other hand, increases the number of sources to separate; (b) are more flexible at multi-channel separation; and (c) can be readily configured to perform dereverberation besides separation \cite{Wang2024USDnet}, while MixIT cannot.
On the other hand, when close-talk mixtures are available, M2M can be readily configured weakly-supervised to leverage the weak supervision afforded by close-talk mixtures.

\subsection{Weakly-Supervised Speech Separation}

SuperM2M, building upon UNSSOR \cite{Wang2023UNSSOR} and M2M \cite{Wang2024M2MSPL}, can be configured to leverage close-talk mixtures as a weak supervision to enhance far-field mixtures.
In the literature, there are earlier studies on weakly-supervised speech enhancement and source separation.
In \cite{Stoller2018, Zhang2018WeaklySupervisedASS}, discriminators, essentially source prior models trained in an adversarial way, are used to help separation models produce separation results with distributions close to clean sources.
In \cite{Chang2019MIMOSpeech}, separation models are jointly trained with ASR models to leverage the weak supervision of word transcriptions.
In \cite{Pishdadian2020FindStrength}, a pre-trained sound classifier is employed to check whether separated signals can be classified into target sound classes, thereby promoting separation.
These approaches require clean sources, human annotations (e.g., word transcriptions), and source prior models (e.g., discriminators, ASR models, and sound classifiers).
In comparison, M2M needs close-talk and far-field mixture pairs, which can be obtained during data collection by using close-talk in addition to far-field microphones, and it does not require source prior models.
In addition, close-talk mixtures exploited in M2M can provide a \textit{sample-level} supervision, offering much more fine-grained supervision than source prior models, word transcriptions, and segment-level sound class labels.

\section{Problem Formulation}\label{proposed_problem_formulation}

We start with describing the hypothesized physical models, then propose to formulate speech enhancement as a blind deconvolution problem, and at last overview the proposed algorithm for speech enhancement.

\ZQHL{
\subsection{Physical Model}\label{physical_model}

In noisy-reverberant conditions with a compact far-field $P$-microphone array and a single target speaker wearing a close-talk microphone,
the physical model for each real-recorded mixture can be formulated, in the short-time Fourier transform (STFT) domain, as follows.

At a designated reference far-field microphone $q \in \{1,\dots,P\}$, the real-recorded mixture is formulated as
\begin{align} 
Y_q(t,f) &= X_q(t,f) + V_q(t,f), \label{phymodel_freq_ff_ref}
\end{align}
where $t$ indexes $T$ frames, $f$ indexes $F$ frequency bins, and $Y_q(t,f)$, $X_q(t,f)$, and $V_q(t,f)$ respectively denote the STFT coefficients of the mixture, speaker image of the target speaker, and non-speech signals at time $t$, frequency $f$ and microphone $q$.
In the rest of this paper, when dropping indices $t$ and $f$, we refer to the corresponding spectrograms.
We emphasize that $V_q$ could contain multiple strong, non-stationary directional as well as diffuse noises.
In this paper, we model them using a single, combined source.

Let us denote the index of the close-talk microphone as $0$, which is different from the set of indices $\{1,\dots,P\}$ used for far-field microphones.
This way, we can index any of the non-reference microphones\footnote{By non-reference microphones, we mean the close-talk microphone and $P-1$ non-reference far-field microphones.} using $p\in \{0,1,\dots,P\}$, with $p\neq q$.
We then formulate the mixture captured at microphone $p$ as
\begin{align} 
Y_p(t,f&) = X_p(t,f) + V_p(t,f) \nonumber \\ 
        &= \mathbf{h}_p(f)^\H \overline{\mathbf{X}}_q(t,f;p) + V_p(t,f) + \varepsilon_p(t,f) \nonumber \\
        &= \mathbf{h}_p(f)^\H \overline{\mathbf{X}}_q(t,f;p) + \mathbf{r}_p(f)^\H \overline{\mathbf{V}}_q(t,f;p) + \varepsilon'_p(t,f).
\label{phymodel_freq_ff_non_ref}
\end{align}
In row $2$, following narrowband approximation \cite{Talmon2009, Gannot2017}, we approximate the speaker image captured at microphone $p$ (i.e., $X_p(\cdot,f)$) as a linear convolution between the speaker image captured at microphone $q$ (i.e., $X_q(\cdot,f)$) and a linear filter $\mathbf{h}_p(f)$.
That is, $X_p(t,f)\approx\mathbf{h}_p(f)^\H \overline{\mathbf{X}}_q(t,f;p)$, where $\overline{\mathbf{X}}_q(t,f;p)=[X_q(t-I_p+1,f),\dots,X_q(t+J_p,f)] \in \CC^{I_p+J_p}$ and $\mathbf{h}_p(f)\in \CC^{I_p+J_p}$, and we use $\varepsilon_p$ to denote the modeling error.
Notice that the past and future filter taps, $I_p$ and $J_p$, are dependent on the microphone index (i.e., $p$) of the speaker image $X_p$ we want to approximate.
$\mathbf{h}_p(f)$ can be interpreted as the relative transfer function (RTF) relating the speaker image $X_q$ (captured by the reference far-field microphone $q$) to the speaker image captured at another microphone $p$ (i.e., $X_p$).
In row $3$, we use the same trick to approximate non-speech signals $V_p(t,f)\approx \mathbf{r}_p(f)^\H \overline{\mathbf{V}}_q(t,f;p)$, where $\overline{\mathbf{V}}_q(t,f;p)=[V_q(t-I_p+1,f),\dots,V_q(t+J_p,f)] \in \CC^{I_p+J_p}$ and $\mathbf{r}_p(f)\in \CC^{I_p+J_p}$, and $\varepsilon'_p$ absorbs the incurred modeling error and $\varepsilon_p$.

To facilitate understanding, we provide several comments:
\begin{itemize}[leftmargin=*,noitemsep,topsep=0pt]
\item For simplicity, we use $I_p-1$ past and $J_p$ future taps for both $\overline{\mathbf{X}}_q(t,f;p)$ and $\overline{\mathbf{V}}_q(t,f;p)$, although it might be better to use different taps for different sources.
\item For the filter taps $I_p$ and $J_p$, we have a subscript $p$ to indicate that we can use different number of filter taps for each non-reference microphone $p$.
It may be a good idea to use different filter taps for the non-reference far-field microphones and the close-talk microphone, especially if they are placed on different devices or not synchronized with each other.
\item $\mathbf{r}_p(f)^\H \overline{\mathbf{V}}_q(t,f;p)$ could be a crude approximation of $V_p(t,f)$, as there could be multiple directional and diffuse noise sources in $V$, rather than a single directional source like in $X_q$.
\item In the speaker image captured by the close-talk microphone (i.e., $X_0$), the direct-path signal of the target speaker is often much stronger than its reverberation. Therefore, $X_0$ can be largely viewed as the dry source signal.
In this case, $\mathbf{h}_0(f)$ can be interpreted as a deconvolutional filter \ZQHLRtwo{that can reverse the speaker image $X_q$ back to the speech source signal}.
\end{itemize}

}

\subsection{Formulating Speech Enhancement as Blind Deconvolution}\label{blind_deconv}

As is suggested by our preliminary studies, UNSSOR \cite{Wang2023UNSSOR} and M2M \cite{Wang2024M2MSPL}, both close-talk and far-field mixtures contain weak supervision for speaker separation.
\ZQHLRtwo{They \cite{Wang2023UNSSOR, Wang2024M2MSPL} exploit the weak supervision by formulating speaker separation as solving a blind deconvolution problem.

\begin{figure*}
  \centering
  \includegraphics[width=14.5cm]{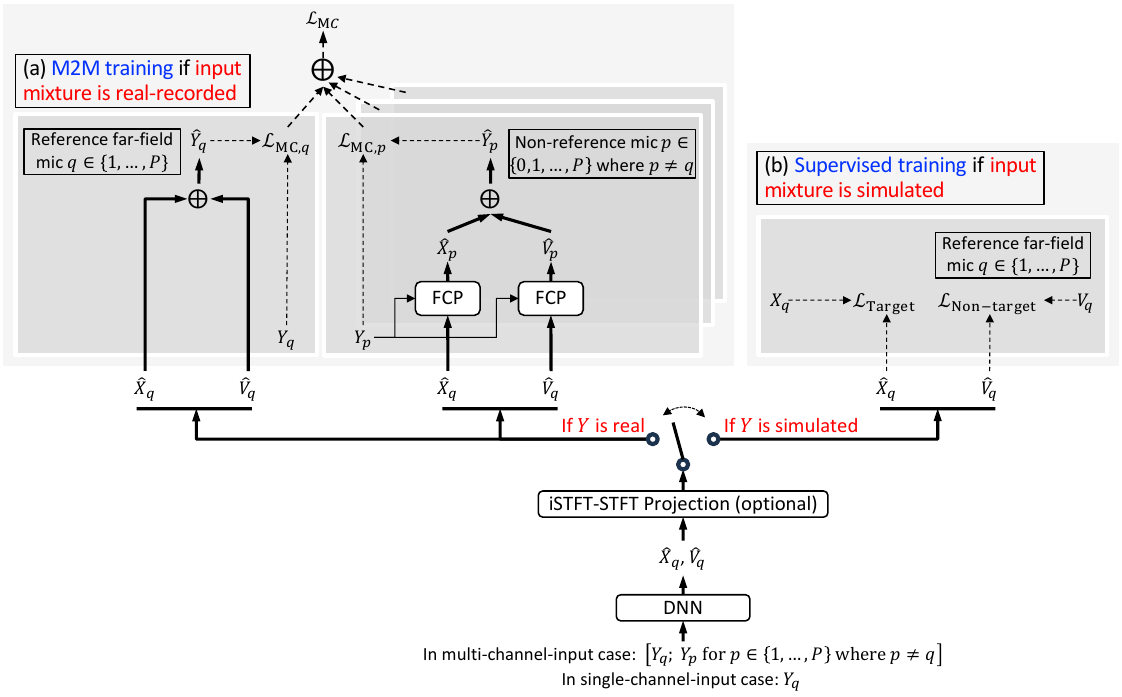} %
  \caption{
  \ZQHL{Illustration of SuperM2M, which consists of (a) M2M training on real close-talk and far-field mixture pairs (described in Section \ref{proposed_M2M}); and (b) supervised training on simulated far-field mixtures (described in Section \ref{proposed_cotraining}).
  M2M trains the DNN to separate far-field mixtures to two sources that can be linearly filtered to reconstruct far-field and close-talk mixtures.
  In SuperM2M, we alternately feed in mini-batches of real mixtures and mini-batches of simulated mixtures, and train the same DNN by alternating between M2M training (if the input mixtures are real) and supervised learning (if the input mixtures are simulated).}
  }
  \label{system_figure}
\end{figure*}

Similar to speaker separation, speech enhancement can be viewed as two-source separation, with one source being speech and the other being all non-speech sources combined.
With this understanding, following UNSSOR \cite{Wang2023UNSSOR} and M2M \cite{Wang2024M2MSPL}, we can also formulate speech enhancement as solving a blind deconvolution problem, and adapt UNSSOR and M2M for speech enhancement.
In detail, the problem, formulated below,} finds sources, $X_q(\cdot,\cdot)$ and $V_q(\cdot,\cdot)$, and filters, $\mathbf{h}_{\cdot}(\cdot)$ and $\mathbf{r}_{\cdot}(\cdot)$, that are most consistent with the physical models in (\ref{phymodel_freq_ff_ref}) and (\ref{phymodel_freq_ff_non_ref}):
\begin{align}\label{ideal_loss}
&\underset{\substack{X_q(\cdot,\cdot),V_q(\cdot,\cdot),\\\mathbf{h}_{\cdot}(\cdot),\mathbf{r}_{\cdot}(\cdot)}}{\text{argmin}}
\Big( \sum\limits_{t,f} \Big| Y_q(t,f) - X_q(t,f) - V_q(t,f) \Big|^2 + \nonumber \\
&\sum\limits_{p=0,p\neq q}^{P} \sum\limits_{t,f} \Big| Y_p(t,f) - \mathbf{h}_p(f)^\H \overline{\mathbf{X}}_q(t,f;p) - \mathbf{r}_p(f)^\H \overline{\mathbf{V}}_q(t,f;p) \Big|^2 \Big),
\end{align}
where $|\cdot|$ computes magnitude.
As is suggested in UNSSOR \cite{Wang2023UNSSOR}, this problem is non-convex and difficult to be solved, since the speech source, noise source, and linear filters are all unknown and need to be estimated.
It is known not solvable if no prior knowledge is assumed about the sources or filters \cite{Levin2011}.
In our preliminary work, UNSSOR \cite{Wang2023UNSSOR}, which models source priors via unsupervised deep learning, is proposed to tackle this category of blind deconvolution problems. It is shown effective at separating reverberant multi-speaker mixtures to reverberant speaker images in simulated conditions.

\subsection{\ZQHLRtwo{Overview of Proposed Algorithms}}\label{proposed_overview}

Building upon the preliminary successes of UNSSOR \cite{Wang2023UNSSOR} and M2M \cite{Wang2024M2MSPL} in speaker separation on simulated data, this paper adapts UNSSOR and M2M training for neural speech enhancement and performs training and evaluation on real-recorded data.

\ZQHLRtwo{
We first give UNSSOR and M2M a review in Section \ref{proposed_M2M}.
We then propose to improve them by addressing potential synchronization issues between close-talk and far-field microphones in Section \ref{estimate_future_taps_description}.
Next, in Section \ref{proposed_cotraining} we point out several weaknesses of UNSSOR and M2M, and propose to address them by combining UNSSOR and M2M training on real data with supervised learning on simulated data.
}

\section{\ZQHLRtwo{M2M Review, and Adapting It for Speech Enhancement}}\label{proposed_M2M}

\ZQHLRtwo{
Since we formulate speech enhancement as two-source separation, we can readily adapt UNSSOR and M2M (originally designed for speaker separation) for speech enhancement.
This section reviews UNSSOR and M2M training, but in the context of adapting them for speech enhancement.
}

Fig. \ref{system_figure}(a) illustrates M2M training for speech enhancement.
The DNN takes in far-field mixtures as input and produces an estimate $\hat{X}_q$ for the target speaker and an estimate $\hat{V}_q$ for non-target signals.
Each estimate is then linearly filtered via forward convolutive prediction \cite{Wang2021FCPjournal} to optimize a so-called mixture-constraint loss \cite{Wang2023UNSSOR}, which encourages the filtered estimates to add up to the close-talk mixture and each far-field mixture, thereby exploiting the weak supervision afforded by close-talk and far-field mixtures for enhancement.
This section describes the DNN setup, loss functions, and FCP filtering.

\subsection{DNN Configurations}\label{DNN_setup}

The DNN is trained to perform complex spectral mapping \cite{Wang2020chime, Wang2020css, Tan2022NSF}, where the real and imaginary (RI) components of far-field mixtures are stacked as input features for the DNN to predict the RI components of $\hat{X}_q$ and $\hat{V}_q$.
The DNN setup is described later in Section \ref{mis_config}, and the loss function next.
We can optionally apply iSTFT-STFT projection to $\hat{X}_q$ and $\hat{V}_q$ before loss computation (see later Section \ref{iSTFT_STFT_proj} for details).

\subsection{Mixture-Constraint Loss}\label{mc_loss}

Following UNSSOR \cite{Wang2023UNSSOR}, M2M \cite{Wang2024M2MSPL} and the objective in (\ref{ideal_loss}), we design the following mixture-constraint (MC) loss, which regularizes the DNN estimates $\hat{X}_q$ and $\hat{V}_q$ to have them respectively approximate $X_q$ and $V_q$, by checking whether they can be utilized to reconstruct the recorded mixtures:
\begin{align}
\mathcal{L}_{\text{MC}} = \mathcal{L}_{\text{MC},q} + \sum_{p=0, p\neq q}^{P} \alpha_p \times  \mathcal{L}_{\text{MC},p},\label{MC_loss}
\end{align}
where the two loss terms respectively follow the ones in (\ref{ideal_loss}) and will be detailed next, and $\alpha_p$, a weighting term, is set to $1.0$ for the close-talk microphone and to $1/(P-1)$ for non-reference far-field microphones. 

$\mathcal{L}_{\text{MC},q}$, following the first term in (\ref{ideal_loss}), is the MC loss at the reference far-field microphone $q$:
\begin{align}
\mathcal{L}_{\text{MC},q} &= \sum\limits_{t,f} \mathcal{F} \Big( Y_q(t,f), \hat{Y}_q(t,f) \Big) \nonumber \\
&=\sum\limits_{t,f} \mathcal{F} \Big( Y_q(t,f), \hat{X}_q(t,f) + \hat{V}_q(t,f) \Big),
\label{MC_loss_ref}
\end{align}
where, in row $2$, the DNN estimates $\hat{X}_q$ and $\hat{V}_q$ are utilized to reconstruct the mixture $Y_q$ via $\hat{Y}_q = \hat{X}_q + \hat{V}_q$, and
$\mathcal{F}(\cdot, \cdot)$, to be described in (\ref{L_D}), is a distance function.

Similarly, $\mathcal{L}_{\text{MC},p}$, following the second term in (\ref{ideal_loss}), is the MC loss at each non-reference microphone $p$:
\begin{align}
&\mathcal{L}_{}{}_{\text{MC},p} = \sum\limits_{t,f} \mathcal{F}\Big( Y_p(t,f), \hat{Y}_p(t,f) \Big) \nonumber \\
&= \sum\limits_{t,f} \mathcal{F}\Big( Y_p(t,f), \hat{X}_p(t,f) + \hat{V}_p(t,f) \Big) \nonumber \\
&=\sum\limits_{t,f} \mathcal{F}\Big( Y_p(t,f), %
\hat{\mathbf{h}}_p(f)^\H \overline{\hat{\mathbf{X}}}_q(t,f;p) + \hat{\mathbf{r}}_p(f)^\H \overline{\hat{\mathbf{V}}}_q(t,f;p)
\Big), \label{MC_loss_nonref}
\end{align}
where $\hat{X}_p(t,f) = \hat{\mathbf{h}}_p(f)^\H \overline{\hat{\mathbf{X}}}_q(t,f;p)$, with $\overline{\hat{\mathbf{X}}}_q(t,f;p)=\big[ \hat{X}_q(t-I_p+1,f),\dots,\hat{X}_q(t+J_p,f)\big]^\T \in \CC^{I_p+J_p}$ and $\hat{\mathbf{h}}_p(f)\in \CC^{I_p+J_p}$, and $\hat{V}_p(t,f)=\hat{\mathbf{r}}_p(f)^\H \overline{\hat{\mathbf{V}}}_q(t,f;p)$, with $\overline{\hat{\mathbf{V}}}_q(t,f;p)=\big[ \hat{V}_q(t-I_p+1,f),\dots,\hat{V}_q(t+J_p,f)\big]^\T \in \CC^{I_p+J_p}$ and $\hat{\mathbf{r}}_p(f) \in \CC^{I_p+J_p}$.
$\hat{\mathbf{h}}_p(f)$ and $\hat{\mathbf{r}}_p(f)$ are both estimated filters to be described in Section \ref{fcp_description}.

Following \cite{Wang2021compensation}, $\mathcal{F}(\cdot, \cdot)$ computes a loss on the estimated RI components and their magnitude:
\begin{align}
\mathcal{F} \Big( Y_r(t,f), \hat{Y}_r(t,f) \Big) = & \frac{1}{\sum\nolimits_{t',f'} |Y_r(t',f')|} \mathcal{G} \Big( Y_r(t,f), \hat{Y}_r(t,f) \Big), \label{L_D} \\
\mathcal{G} \Big( Y_r(t,f), \hat{Y}_r(t,f) \Big) = & \Big| \mathcal{R}(Y_r(t,f)) - \mathcal{R}(\hat{Y}_r(t,f)) \Big| \nonumber \\
&+\Big| \mathcal{I}(Y_r(t,f)) - \mathcal{I}(\hat{Y}_r(t,f)) \Big| \nonumber \\
&+\Big| |Y_r(t,f)| - |\hat{Y}_r(t,f)| \Big|, \label{L_RI_Mag}
\end{align}
where $r\in \{0,1,\dots,P\}$ indexes all the close-talk and far-field microphones, $|\cdot|$ computes magnitude, $\mathcal{R}(\cdot)$ and $\mathcal{I}(\cdot)$ respectively extract RI components, and the denominator in (\ref{L_D}) balances the losses at different microphones and across training mixtures.

Notice that the DNN can use all or a subset of the far-field microphone signals as the input and for loss computation.
For example, a monaural enhancement model can be trained by just using the reference microphone signal as the input but computing the loss on all the microphone signals.

\subsection{FCP for Filter Estimation}\label{fcp_description}

To compute $\mathcal{L}_{\text{MC}}$, we need to first compute the linear filters (i.e., RTFs) in (\ref{MC_loss_nonref}).
Following UNSSOR \cite{Wang2023UNSSOR}, we leverage FCP \cite{Wang2021FCPjournal, Wang2021FCPwaspaa} for filter estimation, based on the DNN estimates and observed mixtures.

Assuming that the target speaker is non-moving within each utterance, we estimate, e.g., the filter $\hat{\mathbf{h}}_p(f)$ in (\ref{MC_loss_nonref}), by solving the following problem:
\begin{align}\label{fcp_proj_mixture}
\hat{\mathbf{h}}_p(f)  =
\underset{\mathbf{h}_p(f)}{\text{argmin}}
\sum\limits_t \frac{\Big| Y_p(t,f) - \mathbf{h}_p(f)^\H \overline{\hat{\mathbf{X}}}_q(t,f;p) \Big|^2}{\hat{\lambda}_p(t,f)},
\end{align}
where $\hat{\lambda}$, to be described in (\ref{FCPweight}), is a weighting term.
The objective in (\ref{fcp_proj_mixture}) is quadratic, where a closed-form solution can be readily computed.
We use the same method in (\ref{fcp_proj_mixture}) (i.e., linearly projecting DNN estimate to observed mixture) to compute all the other filters, and then plug the closed-form solutions to compute the $\mathcal{L}_{\text{MC}}$ loss and train the DNN.

In (\ref{fcp_proj_mixture}), $\hat{\lambda}$ is a weighting term balancing the importance of each T-F unit, as different T-F units usually have diverse energy levels.
Following \cite{Wang2021FCPjournal}, it is defined as
\begin{align}\label{FCPweight}
\hat{\lambda}_r(t,f) = \xi\times \text{max}(|Y_r|^2) + |Y_r(t,f)|^2,
\end{align}
where $r$ indexes all the microphones, $\xi$ (tuned to $10^{-2}$ in this study) floors the weighting term, and $\text{max}(\cdot)$ extracts the maximum value of a power spectrogram.
We compute $\hat{\lambda}$ differently for different microphones, as the energy level of each source can be very different at close-talk and far-field microphones, and deployed microphones, even if placed close to each other, often produce very different gain levels in real-world scenarios.

\section{\ZQHLRtwo{On-the-fly Estimation of Future Taps $J_0$ for Close-Talk Microphone}}\label{estimate_future_taps_description}

The FCP filter taps used in M2M training (i.e., $I_p$ and $J_p$ for $p\in \{0,1,\dots,P\}$ and $p\neq q$) are hyperparameters to tune.
Their ideal values are likely different for different utterances, while it is tricky and cumbersome to tune each one of them individually for each utterance.
For simplicity, in UNSSOR \cite{Wang2023UNSSOR} and M2M \cite{Wang2024M2MSPL}, the value of each filter tap is configured shared for all the training utterances.

This strategy should be improved when dealing with real-recorded data, such as CHiME-4 \cite{Vincent2016CHiME4Analysis}.
In CHiME-4, we observe that the far-field microphones are reasonably synchronized as they are placed on, and processed by, the same device, but the close-talk microphone, placed on a different device, is not accurately synchronized with the far-field microphones.
Their time-misalignment can be as large as $50$ ms.
On the other hand, the distance between the far-field microphone array and the close-talk microphone is unknown and can vary from utterance to utterance.

In this context, for simplicity, for all the training utterances, we set the past filter taps the same for all the non-reference microphones (i.e., all the $I_p$ for $p\in \{0,1,\dots,P\}$ and $p\neq q$ are configured the same) and set the future tap $J_p$ to $1$ for all the non-reference far-field microphones, while we propose to, at each training step, estimate $J_0$ (i.e., the future taps for the close-talk microphone) for each training utterance in the mini-batch by solving the problem below.
The problem below, following the $\mathcal{L}_{\text{MC},0}$ loss in (\ref{MC_loss_nonref}), enumerates a set of future taps and finds the one that leads to the best approximation of the close-talk mixture based on very short FCP filters:
\begin{align}
J_0 &= \underset{J_0'\in\Omega}{\text{argmin}} \sum\limits_{t,f} \mathcal{F}\Big( Y_0(t,f),\hat{\mathbf{h}}_0(f)^\H \vec{\hat{\mathbf{X}}}_q(t,f;0) + \hat{\mathbf{r}}_0(f)^\H \vec{\hat{\mathbf{V}}}_q(t,f;0) \Big),\label{future_tap_estimation}
\end{align}
where $\vec{\hat{\mathbf{X}}}_q(t,f;0)=\big[ \hat{X}_q(t+J_0'-K+1,f),\dots,\hat{X}_q(t+J_0',f)\big]^\T \in \CC^K$, $\vec{\hat{\mathbf{V}}}_q(t,f;0)=\big[ \hat{V}_q(t+J_0'-K+1,f),\dots,\hat{V}_q(t+J_0',f)\big]^\T \in \CC^K$, $K$ is set to a small value ($3$ in this study) so that the filter is short and the amount of computation spent on solving this problem is small, $\Omega=\{0,1,\dots,R\}$ denotes a set of future taps to enumerate (with $R$ tuned to $8$ for CHiME-4 to account for potentially large errors in synchronization), $J_0'\in\Omega$ denotes an enumerated candidate future tap, and $\hat{\mathbf{h}}_0(f)$ and $\hat{\mathbf{r}}_0(f)$ are computed in the same way as in (\ref{fcp_proj_mixture}).

Note that we run (\ref{future_tap_estimation}) at each training step to estimate $J_0$ for each training utterance in the mini-batch.
We stop gradients for the operations in (\ref{future_tap_estimation}) in the forward pass, and hence no back-propagation is performed for the operations in (\ref{future_tap_estimation}).
The estimated $J_0$ is then used for computing the $\mathcal{L}_{\text{MC},0}$ loss in (\ref{MC_loss_nonref}).

\section{SuperM2M}\label{proposed_cotraining}

\ZQHLRtwo{
We first point out the weaknesses of M2M training, and then propose SuperM2M to address the weaknesses.
Next, we discuss the necessity of close-talk mixtures in SuperM2M.
}

\subsection{\ZQHLRtwo{Weaknesses of M2M Training}}\label{m2m_weakness}

M2M is a weakly-supervised speech enhancement algorithm that can learn from the weak supervision afforded by close-talk and far-field mixtures.
It can also be viewed, with a grain of salt, as an unsupervised enhancement algorithm, where the DNN is trained to produce two source estimates that can be linearly filtered to best \textit{explain} (i.e., reconstruct) the close-talk and far-field mixtures.
In this regard, the resulting enhancement system needs to deal with three tricky issues.

First, the source estimates could be permuted randomly.
That is, they could respectively correspond to speech and noise, or the opposite, since the two estimates and their linearly-filtered results are only constrained to sum up to the mixtures.

Second, the source estimates could suffer from frequency permutation \cite{Sawada2019BSSReview}, a common problem that needs to be dealt with in many frequency-domain unsupervised separation algorithms such as independent vector analysis, spatial clustering, and UNSSOR \cite{Wang2023UNSSOR}.
Since FCP is performed in each frequency independently from the others, even though speech and noise sources are accurately separated in each frequency, the separation results of each source at different frequencies are not guaranteed to be grouped into the same output spectrogram.

Third, since, in realistic cases, the noise component $V$ usually consists of an unknown number of directional and diffuse sources, in unsupervised separation the model would lack an idea to produce one estimate exactly corresponding to target speech and the other exactly corresponding to all the noise sources combined.
In other words, sources are ambiguous to the model.
It is possible that, even if one estimate contains the target speech plus some noise sources and the other estimate absorbs the rest noise sources, the mixture-constraint loss can still be very low.
The fundamental causes of this problem are that, in unsupervised setups, (a) the model lacks an exact concept about what the target source should be like; and (b) the hypothesized number sources (in this paper, $2$) is not guaranteed to match the actual number of sources (i.e., speech source plus an unknown number of noise sources) in every training mixture.

These issues do not exist in supervised approaches, as the oracle simulated speech and noise signals used in supervised approaches can penalize the DNN estimates to naturally avoid source and frequency permutation, and resolve source ambiguity.
This motivates us to combine M2M training with supervised learning, leading to SuperM2M, which is described next.

\subsection{Supervised and Mixture-to-Mixture Co-Learning}

The previous subsection points out that M2M suffers from source and frequency permutation, and source ambiguity.
On the other hand, although M2M training can be performed on real mixtures, there may not be many paired close-talk and far-field real-recorded mixtures available, as collecting real data is effort-consuming.
In comparison, supervised models can be readily trained on massive simulated mixtures, as one can easily simulate as many mixtures as one considers sufficient.
In addition, they do not suffer from source and frequency permutation, and source ambiguity.

In this context, we propose to train the same DNN model with both M2M training and supervised learning to combine their strengths.
We name the algorithm SuperM2M.
See Fig. \ref{system_figure} for an illustration, where the supervised learning part is shown in Fig. \ref{system_figure}(b).
Notice that the DNN in M2M training is designed to directly produce target and non-target estimates. 
This makes M2M training capable of being easily integrated with supervised training, where the models are usually designed to directly produce target estimates.

In detail, at each training step, we sample either a mini-batch of real close-talk and far-field mixture pairs or a mini-batch of simulated far-field mixtures for DNN training.
\ZQHL{The loss for the mini-batch of real data is $\mathcal{L}_{\text{MC}}$ in (\ref{MC_loss}), and the loss for the mini-batch of simulated data is}
\begin{align}
\mathcal{L}_{\text{SIMU},q} &= \mathcal{L}_{\text{Target},q} + \mathcal{L}_{\text{Non-target},q}, \label{SIMU_loss} \\
\mathcal{L}_{\text{Target},q} &= \frac{1}{\sum\nolimits_{t,f} |Y_q(t,f)|} \sum_{t,f} \mathcal{G}\Big( X_q(t,f), \hat{X}_q(t,f) \Big), \\ 
\mathcal{L}_{\text{Non-target},q} &= \frac{1}{\sum\nolimits_{t,f} |Y_q(t,f)|}\sum_{t,f} \mathcal{G}\Big( V_q(t,f), \hat{V}_q(t,f) \Big),
\end{align}
where $\mathcal{G}(\cdot,\cdot)$ is defined in (\ref{L_RI_Mag}), $X_q$ and $V_q$ are obtained through simulation, and the denominator balances the loss values with the ones in M2M training.

\begin{figure*}
  \centering  
  \includegraphics[width=18cm]{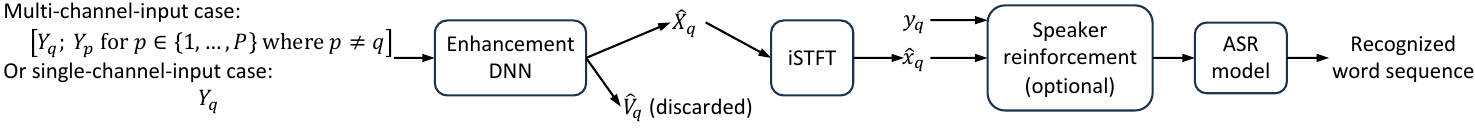}
  \caption{
    Robust ASR pipeline, where enhanced speech $\hat{x}_q = \text{iSTFT}(\hat{X}_q)$ is fed to backend ASR models for recognition.
    No joint training is performed. An optional speaker reinforcement module \cite{Zorila2022SpeakerReinforcement}, which adds a scaled version of the input mixture signal $y_q$ to $\hat{x}_q$, can be included.
  }
  \label{enh_asr_pipeline_figure}
\end{figure*}

\subsection{Necessity of Close-Talk Mixtures}

So far, we hypothesize that, during training, a paired close-talk mixture is always available for far-field mixtures.
It is leveraged as a weak supervision for training by optimizing a mixture-constraint loss (i.e., $\mathcal{L}_{\text{MC},0}$ in (\ref{MC_loss_nonref})) defined on it.

When close-talk mixtures are not available, we find that we can still train enhancement models successfully via SuperM2M, where, in the M2M part, the DNN is trained to only recover far-field mixtures, meaning that M2M training is unsupervised.\footnote{When close-talk mixtures are not available, M2M \cite{Wang2024M2MSPL} regresses to UNSSOR \cite{Wang2023UNSSOR}.
In our paper, we prefer to still call our algorithms SuperM2M, rather than SuperUNSSOR, just to avoid creating too many new names.}
This is a desirable property, as this means that we only need a set of real-recorded far-field multi-channel mixtures (which are easier to record than paired far-field and close-talk mixtures), and together with a set of simulated mixtures, we can train an enhancement system via SuperM2M, which could generalize better to real mixtures than purely-supervised models trained only on the simulated mixtures.

\section{Experimental Setup}\label{experimental_setup}

Our main goal is to show that SuperM2M can generalize better to real data than purely supervised models trained on simulated data.
We follow the robust ASR pipeline in Fig. \ref{enh_asr_pipeline_figure} for evaluation, not using any joint frontend-backend training.

We do not use $\hat{X}_q$ to derive linear beamforming results for ASR \cite{WDLreview, HaebUmbach2019SPM, Haeb-Umbach2020}, although this has been extremely popular, as we would like to validate whether the enhanced speech $\hat{X}_q$ itself is close to target speech and whether $\hat{X}_q$ itself can yield better ASR performance.
We do not jointly train enhancement models with ASR models, as this requires knowledge of ASR models and would not accurately reflect the accuracy of $\hat{X}_q$ itself.
We aim at building enhancement models that can produce enhanced speech with low distortion and high reduction to non-target signals.
This way, the enhancement models could improve the robustness of many subsequent applications not limited to ASR.

In a nutshell, our main goal is to show, through SuperM2M, whether $\hat{X}_q$ itself would be better on real test data.
We validate SuperM2M on CHiME-4 \cite{Vincent2016CHiME4Analysis}, a dataset consisting of simulated mixtures and real-recorded close-talk and far-field mixture pairs.
To further show the effectiveness and potential of SuperM2M, a minor goal is to show whether SuperM2M can lead to state-of-the-art ASR performance on CHiME-4.
The rest of this section describes the CHiME-4 dataset, miscellaneous system configurations, comparison systems, evaluation metrics, and several tricks to improve robust ASR performance.

\subsection{CHiME-4 Dataset}

CHiME-4 \cite{Vincent2016CHiME4Analysis} is a major corpus for evaluating robust ASR and speech enhancement algorithms. 
It is recorded by using a tablet mounted with $6$ microphones, with the second microphone on the rear and the others facing front.
The signals are recorded in four representative environments (including cafeteria, buses, pedestrian areas, and streets), where reverberation and directional, diffuse, transient and non-stationary noises naturally exist.
During data collection, the target speaker hand-holds the tablet in a designated environment, and reads text prompts shown on the screen of the tablet.
The target speaker wears a close-talk microphone so that the close-talk mixture can be recorded at the same time along with far-field mixtures recorded by the microphones on the tablet.
The number of simulated and real-recorded 
utterances is listed in Table \ref{chime4_number_mixtures}.
\ZQHL{It should be noted that, in CHiME-4, the room reverberation is weak, and the major challenge is in how to remove the multi-source non-stationary noise signals.}

In the real data of CHiME-4, we observe synchronization errors between the close-talk microphone and far-field microphone array.
Other issues, such as microphone failures, signal clipping, speaker and array movement, and diverse gain levels even if microphones are placed close to each other, happen frequently.
In real-world products, these are typical problems, which increase the difficulties of speech enhancement and ASR.
They need to be robustly dealt with by frontend enhancement systems.

\begin{table}[]
\scriptsize
\centering
\sisetup{table-format=2.2,round-mode=places,round-precision=2,table-number-alignment = center,detect-weight=true,detect-inline-weight=math}
\caption{Number of utterances in CHiME-4 (all are six-channel)}
\label{chime4_number_mixtures}
\setlength{\tabcolsep}{4pt}
\begin{tabular}{
c %
ccc
}
\toprule
Type & {Training Set} & {Validation Set} & {Test Set} \\
\midrule
{\multirow{2}{*}{SIMU}} & {\multirow{2}{*}{$7,138$ ($\sim$$15.1$ h)}} & \multirow{2}{*}{\begin{tabular}[c]{@{}c@{}}$1,640$ ($\sim$$2.9$ h)\\($410$ in each environ.) \end{tabular}} & \multirow{2}{*}{\begin{tabular}[c]{@{}c@{}}$1,320$ ($\sim$$2.3$ h)\\($330$ in each environ.)\end{tabular}} \\
\\
\midrule
{\multirow{2}{*}{REAL}} & {\multirow{2}{*}{$1,600$ ($\sim$$2.9$ h)}} & \multirow{2}{*}{\begin{tabular}[c]{@{}c@{}}$1,640$ ($\sim$$2.7$ h)\\($410$ in each environ.) \end{tabular}} & \multirow{2}{*}{\begin{tabular}[c]{@{}c@{}}$1,320$ ($\sim$$2.2$ h)\\($330$ in each environ.)\end{tabular}} \\
\\
\bottomrule
\end{tabular}
\end{table}

Depending on the number of microphones that can be used for recognition, there are three official ASR tasks in CHiME-4, including $1$-, $2$- and $6$-channel tasks. In the $1$-channel task, only one of the front microphones can be used for testing; in the $2$-channel task, only two of the front microphones can be used; and in the $6$-channel task, all the six microphones can be used.
For the $1$- and $2$-channel tasks, the microphones that can be used for ASR for each utterance are selected by the challenge organizers to avoid microphone failures.
The selected microphones can vary from utterance to utterance.

\subsection{Evaluation Setup - Robust ASR}

We check whether SuperM2M can improve ASR performance by feeding its enhanced speech to ASR models for decoding, following the pipeline in Fig. \ref{enh_asr_pipeline_figure}.
We consider two ASR models.
\begin{itemize}[leftmargin=*,noitemsep,topsep=0pt]
\item The first one is Whisper Large v2\footnote{\url{https://huggingface.co/openai/whisper-large-v2}} \cite{Radford2023Whisper}, pre-trained on massive data.
We use its text normalizer to normalize hypothesis and reference text before computing WER.
\item The second one is trained on the official CHiME-4 mixtures plus the clean signals in WSJ$0$ by using the public recipe \cite{Chang2022E2EIntegration}\footnote{\url{https://github.com/espnet/espnet/blob/master/egs2/chime4/asr1/conf/tuning/train_asr_transformer_wavlm_lr1e-3_specaug_accum1_preenc128_warmup20k.yaml}} in ESPnet.
It is an encoder-decoder transformer-based model, trained on WavLM features \cite{Chen2022WavLM} and using a transformer language model in decoding.
\ZQHL{It is the current strongest ASR model on CHiME-4.
It is a representative model in leveraging self-supervised learning based representations for ASR and robust ASR \cite{Chang2022E2EIntegration, Zhu2022icassp, Zhu2023TASLP, Du2023TASLP}.}
\end{itemize}
Note that the WERs computed by ESPnet should not be directly compared with the ones by Whisper due to different text normalization.
\ZQHLRtwo{As a side result, to compare the ASR model in ESPnet with the Whisper model, we retrained an ESPnet model by using the same text normalization as the Whisper ASR model. We will report the evaluation results of using this ASR model later in Section \ref{whisper_vs_ESPnet_same_text_norm}.}

\subsection{Evaluation Setup - Speech Enhancement}

We evaluate the enhancement performance of SuperM2M on the simulated test data of CHiME-4.
We consider $1$- and $6$-channel enhancement.
In the $1$-channel case, SuperM2M uses the fifth microphone (CH$5$) signal as input, and the target direct-path signal at CH$5$ is used as the reference for evaluation.
In the $6$-channel case, SuperM2M uses all the microphone signals as input to predict the target speech at CH$5$.

The evaluation metrics include wide-band perceptual evaluation of speech quality (WB-PESQ) \cite{Rix2001}, short-time objective intelligibility (STOI) \cite{H.Taal2011}, signal-to-distortion ratio (SDR) \cite{Vincent2006a}, and scale-invariant SDR (SISDR) \cite{LeRoux2019}.
They are widely-adopted metrics in speech enhancement, which can evaluate the quality, intelligibility, and accuracy of the magnitude and phase of enhanced speech. 

\subsection{Training Setup}

For monaural enhancement, we train SuperM2M using all the $(7,138+1,600)\times 6$ monaural signals.
For $2$-channel enhancement, at each training step we sample $2$ microphones from the front microphones as input, and train the DNN to predict the target speech at the first of the selected microphones.
For $6$-channel enhancement, we train SuperM2M using the $7,138+1,600$ six-channel signals.
The DNN stacks all the six microphones as input to predict the target speech at CH$5$.

For simplicity, we do not filter out microphone signals with any microphone failures in the DNN input and loss.
We would expect SuperM2M to learn to deal with the failures, as it can be trained on real mixtures.

\subsection{Miscellaneous Configurations}\label{mis_config}

For STFT, the window size is $32$ ms, hop size $8$ ms, the square root of Hann window is used as the analysis window, and the synthesis window is designed based on the analysis window to achieve perfect reconstruction.

TF-GridNet \cite{Wang2022GridNetjournal}, which has shown strong separation performance in major benchmarks in supervised speech separation, is used as the DNN architecture.
We consider two setups.
Using the symbols defined in Table I of \cite{Wang2022GridNetjournal}, the first one (denoted as TFGridNetv1) sets its hyper-parameters to $D=100$, $B=4$, $I=2$, $J=2$, $H=200$, $L=4$ and $E=2$, and the second one (denoted as TFGridNetv2) to $D=128$, $B=4$, $I=1$, $J=1$, $H=200$, $L=4$ and $E=4$. Please do not confuse these symbols with the ones in this paper.
The models have $\sim$$6.3$ and $\sim$$5.4$ million parameters respectively.
The v1 model uses \ZQHLRtwo{approximately} half of the computation and memory of v2\footnote{\ZQHLRtwo{
This is because in v1, setting $I=2$ and $J=2$ cuts the sequence length to be modeled by the BLSTMs in TFGridNet by $50\%$, compared with setting $I=1$ and $J=1$ in v2.
In this case, v1 would approximately use half of the computation and memory of v2 for the following reasons.
First, in TFGridNet, the cross-frame self-attention module is much less costly than the sub-band temporal module and the intra-frame full-band module, both of which use BLSTMs, the most computationally-expensive block in each of the two modules.
Second, although the input dimension to the BLSTMs is $200$ in v1 while $128$ in v2, the hidden dimensions of the BLSTMs in v1 and v2 are both set to $200$ and most computation inside BLSTMs is not spent between the input tensor and the hidden tensor.
We highlight that this approximation is rough and is verified only empirically.
}}, and is utilized for faster experimentation.

We train all the enhancement models on $8$-second segments using a mini-batch size of $1$.
At each training step, if the sampled utterance is simulated, we use supervised learning, and if it is real-recorded, we use M2M training.
Adam is employed for optimization.
The learning rate starts from $0.001$ and is halved if the validation loss is not improved in $2$ epochs.

\subsection{Comparison Systems}

We consider the same DNN model trained only on the simulated data of CHiME-4 via supervised learning as the major baseline for comparison.
We use exactly the same configurations as that in SuperM2M for traning.
We denote this baseline as \textbf{Supervised}, to differentiate it with \textbf{SuperM2M}.
Since CHiME-4 is a public and popular dataset, many existing models can be used directly for comparison.

\subsection{Tricks to Improve Robust ASR Performance}\label{tricks_ASR}

To show the effectiveness of SuperM2M, we also check whether it can lead to state-of-the-art ASR performance on CHiME-4.
This subsection describes several tricks that are known to improve robust ASR performance and are commonly used in existing studies.

\subsubsection{SNR Augmentation for Simulated Training Mixtures}\label{lower_SNR}

At each training step, we optionally modify the SNR of the target speech in the CHiME-4 simulated training mixture, on the fly, by $u$ dB, with $u$ uniformly sampled from the range $[-10,+5]$ dB.
In our experiments, we find this technique often producing slightly better enhancement and ASR, but not critical.
Note that we do not change the combinations of speech and noise files to create new mixtures, and we just change the SNR of the existing mixtures in CHiME-4.
No other data augmentation is used for enhancement.

\begin{table*}[t]
\scriptsize
\centering
\captionsetup{justification=centering}
\sisetup{table-format=2.2,round-mode=places,round-precision=2,table-number-alignment=center,detect-weight=true,detect-inline-weight=math}
\caption{SuperM2M vs. purely-supervised models on CHiME-4\\(\#input mics: $6$; ASR model: Whisper Large v2)}
\label{results_SuperM2M_6ch}
\setlength{\tabcolsep}{1.5pt}
\resizebox{1.7\columnwidth}{!}{ %
\begin{tabular}{
l %
c %
c %
c %
c %
c %
c %
c %
S[table-format=2,round-precision=0] %
S[table-format=2.1,round-precision=1] %
S[table-format=2.1,round-precision=1] %
S[table-format=1.2,round-precision=2] %
S[table-format=1.3,round-precision=3] %
S[table-format=2.2,round-precision=2] %
S[table-format=2.2,round-precision=2] %
S[table-format=2.2,round-precision=2] %
S[table-format=2.2,round-precision=2] %
}
\toprule
& & & & & \multirow{2}{*}[-12pt]{\begin{tabular}[c]{@{}c@{}}\#mics\\in\\$\mathcal{L}_{\text{MC}}$\end{tabular}} & & \multirow{3}{*}[-12pt]{\begin{tabular}[c]{@{}c@{}}iSTFT-\\STFT\\proj.?\end{tabular}} & {\multirow{3}{*}[-12pt]{\begin{tabular}[c]{@{}c@{}}Spk.\\reinf.\\$\gamma$ (dB)\end{tabular}}} & \multicolumn{4}{c}{SIMU Test Set (CH5)} & \multicolumn{4}{c}{Official CHiME-4 Test Utterances} \\

\cmidrule(lr){10-13}\cmidrule(lr){14-17}

& & \multirow{2}{*}[-6pt]{\begin{tabular}[c]{@{}c@{}}Training\\data\end{tabular}} & &  & & \multirow{2}{*}[-6pt]{\begin{tabular}[c]{@{}c@{}}SNR\\aug.\end{tabular}} & & & {\multirow{2}{*}[-6pt]{\begin{tabular}[c]{@{}c@{}}SISDR\\(dB)$\uparrow$ \end{tabular}}} & {{\multirow{2}{*}[-6pt]{\begin{tabular}[c]{@{}c@{}}SDR\\(dB)$\uparrow$ \end{tabular}}}} & {{\multirow{2}{*}[-6pt]{\begin{tabular}[c]{@{}c@{}}WB-\\PESQ$\uparrow$ \end{tabular}}}} & & \multicolumn{2}{c}{Val. WER (\%)$\downarrow$} & \multicolumn{2}{c}{Test WER (\%)$\downarrow$} \\

\cmidrule(lr){14-15} \cmidrule(lr){16-17}

{\multirow{1}{*}[0pt]{\rotatebox[origin=c]{0}{System}}} & {\multirow{1}{*}{Approach}} & & DNN arch. & $J_0$ & & & & & & & & {\multirow{1}{*}{STOI$\uparrow$}} & {SIMU} & {REAL} & {SIMU} & {REAL} \\

\midrule

0 & Mixture & - & - & \multicolumn{1}{c}{-} & - & - & - & {-} & 7.506217692752905 & 7.539180193284881 & 1.273397988803459 & 0.8702431371034524 & 7.429816069699903 & 4.957444233794522 & 10.96980976013234 & 7.693500853109973 \\

\midrule

1a & Supervised & S & TFGridNetv1 & - & - & \xmark & \xmark & {-} & 22.233853567730296 & 22.578360742895118 & 3.0811853139689473 & 0.9843909006036486 & 3.5454985479186836 & 28.901617522488 & 3.9340363937138125 & 53.23406235458352 \\
1b & Supervised & S & TFGridNetv2 & - & - & \cmark & \cmark & {-} & \bfseries 22.815442327297095 & \bfseries 23.201049828660896 & \bfseries 3.364026530403079 & \bfseries 0.9877327348654509 & 3.432558889964505 & 58.15013512968415 & 3.789288668320926 & 79.07553901039243 \\

\midrule

2a & Supervised & S & iNeuBe \cite{Wang2020chime} & - & - & {-} & {-} & {-} & 22.0 & 22.4 & {-} & 0.986 & {-} & {-} & {-} & {-} \\
2b & Supervised & S & SpatialNet \cite{Quan2024SpatialNet} & - & - & {-} & {-} & {-} & 22.1 & 22.3 & 2.88 & 0.983 & {-} & {-} & {-} & {-} \\
2c & Supervised & S & USES \cite{Zhang2023USES} & - & - & {-} & {-} & {-} & {-} & 20.6 & 3.16 & 0.983 & {-} & {-} & 4.2 & 78.1 \\
2d & Supervised & S & USES2 \cite{Zhang2024USES2} & - & - & - & {-} & {-} & {-} & 18.8 & 2.94 & 0.979 & {-} & {-} & 4.6 & 12.1 \\

\midrule

3 & SuperM2M & S+R & TFGridNetv1 & {-} & 6 & \xmark & \xmark & {-} & 22.33659448361758 & 22.63268551352279 & 3.1198839563311953 & 0.9844513335492118 & 3.5172636334301384 & 3.928866832092639 & 3.928866832092639 & 4.462023680264723 \\

\midrule

4a & SuperM2M & S+R & TFGridNetv1 & {1} & 6+1 & \xmark & \xmark & {-} & 22.264602772665746 & 22.519353289018003 & 3.0597627621708496 & 0.9843378616620783 & 3.5737334624072283 & 3.8885079262635633 & 3.970223325062035 & 4.038053875187426 \\
4b & SuperM2M & S+R & TFGridNetv1 & {2} & 6+1 & \xmark & \xmark & {-} & 22.247949108481407 & 22.56037238855136 & 3.1112483058914995 & 0.9837983261894265 & 3.505162955792191 & 3.844136985196241 & 4.120140612076096 & 4.162142598624683 \\
4c & SuperM2M & S+R & TFGridNetv1 & {3} & 6+1 & \xmark & \xmark & {-} & 22.179774923758075 & 22.46309740051684 & 3.068768288782149 & 0.9841206415536626 & 3.56163278476928 & 3.8844742043483524 & 4.052936311000828 & 4.08975750995295 \\
4d & SuperM2M & S+R & TFGridNetv1 & {4} & 6+1 & \xmark & \xmark & {-} & 22.3803245691639 & 22.710044684176484 & 3.109135566155116 & 0.9854169526051932 & 3.4809616005162956 & 3.973216086482998 & 3.9805624483043838 & 4.0690760560467405 \\
4e & SuperM2M & S+R & TFGridNetv1 & {5} & 6+1 & \xmark & \xmark & {-} & 22.278870459758874 & 22.60010485681361 & 3.07064790707646 & 0.984785720375682 & 3.5172636334301384 & 3.9651486426525753 & 4.052936311000828 & 4.156972235148131 \\
4f & SuperM2M & S+R & TFGridNetv1 & {6} & 6+1 & \xmark & \xmark & {-} & 22.339721592086732 & 22.69235954266679 & 3.101081421429461 & 0.9847961302987215 & 3.6060019361084223 & 3.94498003307652 & 3.9753928866832093 & 4.213846233390208 \\
4g & SuperM2M & S+R & TFGridNetv1 & {7} & 6+1 & \xmark & \xmark & {-} & 22.205550543286584 & 22.52303918778479 & 3.0961382215673274 & 0.9843291442908499 & 3.5293643110680866 & 3.9933846960590538 & 4.006410256410256 & 4.244868414249522 \\
4h & SuperM2M & S+R & TFGridNetv1 & {8} & 6+1 & \xmark & \xmark & {-} & 22.356131915793274 & 22.651026868914457 & 3.1357882496082423 & 0.9849457777117853 & 3.50919651500484 & 3.928845145415675 & 3.96505376344086 & 4.193164779483998 \\

\midrule

5 & SuperM2M & S+R & TFGridNetv1 & {est.} & 6+1 & \xmark & \xmark & {-} & 22.293973675279908 & 22.595431731068068 & 3.0690434804468443 & 0.9845595854759097 & 3.5898676992578253 & 3.9167439796700414 & 4.083953680727874 & \bfseries 3.965668786515692 \\

\midrule

6a & UNSSOR \cite{Wang2023UNSSOR} & R & TFGridNetv1 & {-} & 6 & \xmark & \xmark & {-} & 7.74501127728515 & 8.11060401569821 & 1.2998775058623515 & 0.8702138218447808 & 4.824136818328493 & 4.158767294582712 & 6.20347394540943 & 4.839460214053048 \\
6b & UNSSOR \cite{Wang2023UNSSOR} & S+R & TFGridNetv1 & {-} & 6 & \xmark & \xmark & {-} & 4.450881271125664 & 4.625870753923576 & 1.323821939031283 & 0.8814539370984228 & 4.808002581477896 & 4.142632406921867 & 5.417700578990901 & 4.875652758388915 \\
[0.5ex]\hdashline\noalign{\vskip 0.5ex}
6c & M2M \cite{Wang2024M2MSPL} & R & TFGridNetv1 & {est.} & 6+1 & \xmark & \xmark & {-} & 9.81428370127017 & 10.312074761824816 & 1.495577291466973 & 0.9016608568481626 & 4.9411100354953215 & 3.953047476906942 & 6.756617038875104 & 4.322423866397808 \\
6d & M2M \cite{Wang2024M2MSPL} & S+R & TFGridNetv1 & {est.} & 6+1 & \xmark & \xmark & {-} & 4.5049174125376865 & 4.636466092929681 & 1.3304595669110617 & 0.8817906046455538 & 4.630525976121329 & 3.949013754991731 & 5.38151364764268 & 4.322423866397808 \\

\midrule

7a & SuperM2M & S+R & TFGridNetv2 & {est.} & 6+1 & \xmark & \xmark & {-} & 22.548556464188028 & 22.87266131204777 & 3.1529978045008398 & 0.9851745435091609 & 3.505162955792191 & 3.900609092009197 & 3.866832092638544 & 4.141461144718474 \\

7b & SuperM2M & S+R & TFGridNetv2 & {est.} & 6+1 & \xmark & \cmark & {-} & 22.70642153512348 & 23.159114919114664 & 3.237787585457166 & 0.986333159671422 & 3.4285253307518553 & 3.8360695413658186 & 3.866832092638544 & 3.981179876945349  \\

7c & SuperM2M & S+R & TFGridNetv2 & {est.} & 6+1 & \cmark & \cmark & {-} & \bfseries 22.849200136282228 & \bfseries 23.214971050443765 & 3.215787200223316 & 0.9867863526257546 & \bfseries 3.384156179412714 & \bfseries 3.8078334879593405 & 3.773779983457403 & 4.043224238663978 \\
[0.5ex]\hdashline\noalign{\vskip 0.5ex}
8a & SuperM2M & S+R & TFGridNetv2 & {est.} & 6+1 & \cmark & \cmark & 10 & {-} & {-} & {-} & {-} & 3.6140690545337204 & 3.8481707071114513 & 3.840984284532672 & 4.244868414249522 \\ 
8b & SuperM2M & S+R & TFGridNetv2 & {est.} & 6+1 & \cmark & \cmark & 15 & {-} & {-} & {-} & {-} & 3.593901258470474 & 3.872373038602719 & \bfseries 3.742762613730356 & 4.105268600382607 \\ 
8c & SuperM2M & S+R & TFGridNetv2 & {est.} & 6+1 & \cmark & \cmark & 20 & {-} & {-} & {-} & {-} & 3.537431429493385 & 3.8643055947722966 & \bfseries 3.737593052109181 & 4.100098236906055 \\ 

\midrule

\rowcolor{shadecolor}
9 & \begin{tabular}{@{}c@{}}Close-talk\\Mixture\end{tabular} & - & - & {-} & {-} & - & - & {-} & {-} & {-} & {-} & {-} & {-} & 3.755395103061595 & {-} & 3.9139651517501677 \\

\bottomrule
\end{tabular}
}
\end{table*}

\subsubsection{iSTFT-STFT Projection}\label{iSTFT_STFT_proj}

We apply inverse STFT (iSTFT) followed by STFT operations to the DNN estimates $\hat{X}_q$ and $\hat{V}_q$ before loss computation, i.e., $\hat{X}_q\coloneqq\text{STFT}(\text{iSTFT}(\hat{X}_q))$ and $\hat{V}_q\coloneqq\text{STFT}(\text{iSTFT}(\hat{V}_q))$.
See Fig. \ref{system_figure} for an illustration.
This often yields slight improvement, as the losses now penalize the RI components and magnitudes extracted from re-synthesized signals, which are the final system output used for human hearing and downstream tasks \cite{Wisdom2018MixtureConsistency,Pandey2019,Wang2021compensation}. Notice that, in Fig. \ref{enh_asr_pipeline_figure}, ASR features are extracted from re-synthesized signals.

\subsubsection{Run-Time Speaker Reinforcement for Robust ASR}\label{spk_reinfocement}

At run time, in default we feed $\hat{x}_q$ for ASR.
Alternatively, we employ a technique named \textit{speaker reinforcement} \cite{Zorila2022SpeakerReinforcement}, where $\hat{x}_q$ is re-mixed with the input mixture $y_q$ at an energy level of $\gamma$ dB before recognition.
See Fig. \ref{enh_asr_pipeline_figure} for an illustration.
That is, $\hat{x}_q+\eta\times y_q$, where $\eta\in \RR_{>0}$ and $\gamma=10\times\text{log}_{10}(\|\hat{x}_q\|^2/\|\eta \times y_q\|^2)$.
We find this technique usually effective for ASR, as the re-mixed input mixture can alleviate distortion to target speech.

\section{Evaluation Results}

This section reports our evaluations results on CHiME-4.
Following earlier studies \cite{Wang2023UNSSOR,Wang2024M2MSPL}, \ZQHL{we set $I_p$ to $20$ for all the non-reference far-field microphones and the close-talk microphone, and $J_p$ to $1$ for all the non-reference far-field microphones.
$J_0$, the future filter tap for the close-talk microphone, can be tuned to a value shared by all the training mixtures, or it can be estimated, on the fly, for each training mixture by using the method described in Section \ref{estimate_future_taps_description}.}
All the six far-field microphones are used for computing $\mathcal{L}_{\text{MC}}$ in (\ref{MC_loss}).
We emphasize that we spent minimal amount of effort on hyper-parameter tuning.
Other hyper-parameter setup could lead to better performance.

\begin{table*}[t]
\scriptsize
\centering
\captionsetup{justification=centering}
\sisetup{table-format=2.2,round-mode=places,round-precision=2,table-number-alignment = center,detect-weight=true,detect-inline-weight=math}
\caption{SuperM2M vs. purely-supervised models on CHiME-4\\(\#input mics: $1$; ASR model: Whisper Large v2)}
\label{results_SuperM2M_1ch}
\setlength{\tabcolsep}{1.5pt}
\resizebox{1.75\columnwidth}{!}{ %
\begin{tabular}{
l %
c %
c %
c %
c %
c %
c %
c %
S[table-format=2,round-precision=0] %
S[table-format=2.1,round-precision=1] %
S[table-format=2.1,round-precision=1] %
S[table-format=1.2,round-precision=2] %
S[table-format=1.3,round-precision=3] %
S[table-format=2.2,round-precision=2] %
S[table-format=2.2,round-precision=2] %
S[table-format=2.2,round-precision=2] %
S[table-format=2.2,round-precision=2] %
}
\toprule
& & & & & \multirow{2}{*}[-12pt]{\begin{tabular}[c]{@{}c@{}}\#mics\\in\\$\mathcal{L}_{\text{MC}}$\end{tabular}} & & \multirow{3}{*}[-12pt]{\begin{tabular}[c]{@{}c@{}}iSTFT-\\STFT\\proj.\end{tabular}} & {\multirow{3}{*}[-12pt]{\begin{tabular}[c]{@{}c@{}}Spk.\\reinf.\\$\gamma$ (dB)\end{tabular}}} & \multicolumn{4}{c}{SIMU Test Set (CH5)} & \multicolumn{4}{c}{Official CHiME-4 Test Utterances} \\

\cmidrule(lr){10-13}\cmidrule(lr){14-17}

& & \multirow{2}{*}[-6pt]{\begin{tabular}[c]{@{}c@{}}Training\\data\end{tabular}} & & & & \multirow{2}{*}[-6pt]{\begin{tabular}[c]{@{}c@{}}SNR\\aug.\end{tabular}} & & & {\multirow{2}{*}[-6pt]{\begin{tabular}[c]{@{}c@{}}SISDR\\(dB)$\uparrow$ \end{tabular}}} & {{\multirow{2}{*}[-6pt]{\begin{tabular}[c]{@{}c@{}}SDR\\(dB)$\uparrow$ \end{tabular}}}} & {{\multirow{2}{*}[-6pt]{\begin{tabular}[c]{@{}c@{}}WB-\\PESQ$\uparrow$ \end{tabular}}}} & & \multicolumn{2}{c}{Val. WER (\%)$\downarrow$} & \multicolumn{2}{c}{Test WER (\%)$\downarrow$} \\

\cmidrule(lr){14-15} \cmidrule(lr){16-17}

{\multirow{1}{*}[0pt]{\rotatebox[origin=c]{0}{System}}} & {\multirow{1}{*}{Approach}} & & DNN arch. & $J_0$ & & & & & & & & {\multirow{1}{*}{STOI$\uparrow$}} & {SIMU} & {REAL} & {SIMU} & {REAL} \\

\midrule

0 & Mixture & - & - & \multicolumn{1}{c}{-} & - & - & - & {-} & 7.506217692752905 & 7.539180193284881 & 1.273397988803459 & 0.8702431371034524 & 7.429816069699903 & 4.957444233794522 & 10.96980976013234 & 7.693500853109973 \\

\midrule

1a & Supervised & S & TFGridNetv1 & - & - & \xmark & \xmark & {-} & 17.052721159927774 & 17.502739632426994 & 2.4357493019465246 & 0.9598704994042264 & 7.139399806389157 & 5.3325803719091605 & 13.156534325889165 & 10.195956775761337 \\
1b & Supervised & S & TFGridNetv2 & - & - & \cmark & \cmark & {-} & 17.075529745808154 & 17.542808687126463 & 2.438363405791196 & 0.9610817692312421 & 7.579057760567926 & 5.19140010487677 & 12.437965260545905 & 9.032383925087787 \\

\midrule

2 & Supervised & S & iNeuBe \cite{Wang2020chime} & - & - & - & {-} & {-} & 15.1 & {-} & {-} & 0.954 & {-} & {-} & {-} & {-} \\

\midrule

3 & SuperM2M & S+R & TFGridNetv1 & {-} & 6 & \xmark & \xmark & {-} & 16.794222991574895 & 17.320037313309793 & 2.4016607089476154 & 0.9602112940675044 & 7.12729912875121 & 5.171231495300714 & 11.796939619520264 & 7.021353601158162 \\

\midrule

4a & SuperM2M & S+R & TFGridNetv1 & {1} & 6+1 & \xmark & \xmark & {-} & 16.84082310624195 & 17.39896686862357 & 2.3754322864792563 & 0.9606356166003033 & 7.103097773475314 & 5.046186115929168 & 11.869313482216707 & 6.928287058580218 \\
4b & SuperM2M & S+R & TFGridNetv1 & {2} & 6+1 & \xmark & \xmark & {-} & 16.863653219558977 & 17.465938016534164 & 2.4458172654563732 & 0.9623696148171832 & 6.990158115521136 & 5.19140010487677 & 11.672870140612076 & 6.8662426968615895 \\
4c & SuperM2M & S+R & TFGridNetv1 & {3} & 6+1 & \xmark & \xmark & {-} & 16.6138287046642 & 17.507949036846473 & 2.4701889178969645 & 0.9605781898099134 & 7.022426589222331 & 4.864668629744665 & 12.257030603804797 & 6.855901969908484 \\
4d & SuperM2M & S+R & TFGridNetv1 & {4} & 6+1 & \xmark & \xmark & {-} & 16.83264474119201 & 17.48211121882364 & 2.4822840224612843 & \bfseries 0.9625962703738554 & 7.103097773475314 & 4.973579121455367 & 11.74524400330852 & 6.871413060338141 \\
4e & SuperM2M & S+R & TFGridNetv1 & {5} & 6+1 & \xmark & \xmark & {-} & 16.90104855326089 & 17.357574660580045 & 2.362516766335025 & 0.9594853026858902 & 7.34107776702162 & 5.288209430841838 & 12.22601323407775 & 7.507367767954087 \\
4f & SuperM2M & S+R & TFGridNetv1 & {6} & 6+1 & \xmark & \xmark & {-} & 16.85934066682151 & 17.37920597843647 & 2.414184342550509 & 0.9616525814351394 & 7.34107776702162 & 5.288209430841838 & 12.22601323407775 & 7.507367767954087 \\ 
4g & SuperM2M & S+R & TFGridNetv1 & {7} & 6+1 & \xmark & \xmark & {-} & 16.963124431443937 & 17.39834809137417 & 2.4209273453011657 & 0.9612934612076076 & 7.228138109067441 & 5.021983784437901 & 12.194995864350703 & 6.902435241197456 \\
4h & SuperM2M & S+R & TFGridNetv1 & {8} & 6+1 & \xmark & \xmark & {-} & 16.62787760161992 & 17.39725872183912 & 2.414387100934982 & 0.9608305905880339 & 7.260406582768635 & 5.026017506353112 & 12.045078577336641 & 6.95413887596298 \\

\midrule

5 & SuperM2M & S+R & TFGridNetv1 & {est.} & 6+1 & \xmark & \xmark & {-} & 16.379974381038636 & 17.44424245604028 & \bfseries 2.5604560643434526 & 0.9621707405031134 & 7.010325911584382 & 4.868702351659877 & 11.693548387096774 & 6.514657980456026 \\

\midrule

6a & UNSSOR \cite{Wang2023UNSSOR} & R & TFGridNetv1 & {-} & 6 & \xmark & \xmark & {-} & 10.259266763361115 & 11.040342632487043 & 1.4179918163653575 & 0.8976332594708465 & 11.023717328170378 & 6.236134080916462 & 15.038254755996691 & 10.423452768729643 \\
6b & UNSSOR \cite{Wang2023UNSSOR} & S+R & TFGridNetv1 & {-} & 6 & \xmark & \xmark & {-} & 11.57728153015628 & 11.708783872236253 & 1.7027648353215419 & 0.9366048267149126 & 7.708131655372701 & 5.239804767859304 & 11.853804797353185 & 8.11747065818727 \\
[0.5ex]\hdashline\noalign{\vskip 0.5ex}
6c & M2M \cite{Wang2024M2MSPL} & R & TFGridNetv1 & {est.} & 6+1 & \xmark & \xmark & {-} & 11.615329997832745 & 12.261265233246604 & 1.7685003023255954 & 0.923511441688311 & 10.717166828009034 & 5.949739824936469 & 14.976220016542596 & 8.60865518845975 \\
6d & M2M \cite{Wang2024M2MSPL} & S+R & TFGridNetv1 & {est.} & 6+1 & \xmark & \xmark & {-} & 10.263214624402197 & 10.411541664128352 & 1.7708911030581502 & 0.9416878780412874 & 7.696030977734754 & 5.659311847041265 & 11.941687344913152 & 8.003722661703119 \\

\midrule

7a & SuperM2M & S+R & TFGridNetv2 & {est.} & 6+1 & \xmark & \xmark & {-} & \bfseries 17.28795669160106 & \bfseries 17.66611348596171 & 2.4430896443851067 & \bfseries 0.963246019069485 & 6.7763794772507255 & 4.933241902303255 & 11.424731182795698 & 6.30784344139393 \\

7b & SuperM2M & S+R & TFGridNetv2 & {est.} & 6+1 & \xmark & \cmark & {-} & 17.101209408406056 & 17.50549155778581 & 2.42846751601407 & 0.9619635709485137 & 7.022426589222331 & 4.7396232503731195 & 11.24379652605459 & 6.230287989245644 \\

7c & SuperM2M & S+R & TFGridNetv2 & {est.} & 6+1 & \cmark & \cmark & {-} & 16.60562385325179 & 17.418666160478683 & 2.510620386221192 & \bfseries 0.9626600245742095 & 7.046627944498225 & 4.783994191440442 & 11.342018196856907 & 6.023473450183548 \\
[0.5ex]\hdashline\noalign{\vskip 0.5ex}
8a & SuperM2M & S+R & TFGridNetv2 & {est.} & 6+1 & \cmark & \cmark & 10 & {-} & {-} & {-} & {-} & \bfseries 5.901097128105841 & \bfseries 4.420959219071437 & \bfseries 8.798593879239041 & \bfseries 5.801147820691795 \\
8b & SuperM2M & S+R & TFGridNetv2 & {est.} & 6+1 & \cmark & \cmark & 15 & {-} & {-} & {-} & {-} & 6.219748305905131 & 4.469363882053972 & 9.336228287841191 & 5.759784912879375 \\
8c & SuperM2M & S+R & TFGridNetv2 & {est.} & 6+1 & \cmark & \cmark & 20 & {-} & {-} & {-} & {-} & 6.397224911261698 & 4.602476705255939 & 9.868693134822166 & 5.858021818933871 \\

\midrule

\rowcolor{shadecolor}
9 & \begin{tabular}{@{}c@{}}Close-talk\\Mixture\end{tabular} & - & - & {-} & {-} & {-} & - & {-} & {-} & {-} & {-} & {-} & {-} & 3.755395103061595 & {-} & 3.9139651517501677 \\

\bottomrule
\end{tabular}
}
\end{table*}

\subsection{SuperM2M vs. Purely-Supervised Models}

Table \ref{results_SuperM2M_6ch} and \ref{results_SuperM2M_1ch} respectively report $6$- and $1$-channel enhancement performance on the fifth microphone of the CHiME-4 simulated test data, and robust ASR performance on the official CHiME-4 test utterances (by using the Whisper Large v2 model for recognition).
The two tables consist of exactly the same set of experiments, and differ only in the number of input microphones to the enhancement models.

On the simulated test data, purely supervised models (in system $1$a-$1$b) trained on the official simulated training data (denoted as \textbf{S} in the ``Training data'' column) produce large improvement over unprocessed mixtures (e.g., in system 1a of Table \ref{results_SuperM2M_6ch}, $22.2$ vs. $7.5$ dB SISDR, and in 1a of Table \ref{results_SuperM2M_1ch}, $17.1$ vs. $7.5$ dB SISDR), and achieve enhancement performance better than existing supervised models such as iNeuBe (based on TCN-DenseUNet) \cite{Wang2020chime, Lu2022}, SpatialNet \cite{Quan2024SpatialNet}, USES \cite{Zhang2023USES} and USES2 \cite{Zhang2024USES2} in both $1$- and $6$-channel cases.
However, the ASR performance is much worse than directly using unprocessed mixtures for ASR, especially in multi-channel cases (e.g., in system $1$a and $0$ of Table \ref{results_SuperM2M_6ch}, $53.23$\% vs. $7.69$\% WER on REAL test data, and in system $1$a and $0$ of Table \ref{results_SuperM2M_1ch}, $10.20$\% vs. $7.69$\% WER on REAL test data).
This degradation has been widely observed by existing studies \cite{HaebUmbach2019SPM, Haeb-Umbach2020}, largely due to the mismatches between simulated training and real-recorded test conditions, and the speech distortion incurred by enhancement.

In system $7$c of Table \ref{results_SuperM2M_6ch} and $7$b of Table \ref{results_SuperM2M_1ch}, we respectively show the results of our best performing SuperM2M models (without employing the speaker reinforcement technique described in Section \ref{spk_reinfocement}) for $6$- and $1$-channel cases.
By training on the official simulated and real data combined (denoted as \textbf{S+R}), SuperM2M obtains clearly better ASR results on the real test data than the purely-supervised models (in system $1$a-$1$b) and unprocessed mixtures (in system $0$), and the enhancement performance on the simulated test data remains strong.
These results indicate that SuperM2M can effectively learn from real data, has better generalizability to real data, and can perform enhancement with low distortion to target speech and high reduction to non-target signals.

Next, we present some ablations results of SuperM2M.

\subsection{Effects of Including Loss on Close-Talk Mixture in $\mathcal{L}_{\text{MC}}$}

First, \ZQHLRtwo{in system $3$}, we report the results of not including close-talk microphone in the loss function.
That is, in (\ref{MC_loss}) we do not include the $\mathcal{L}_{\text{MC},0}$ loss, when training SuperM2M.
We observe clear improvement in ASR performance on the real test data over the purely-supervised model in system $1$a (i.e., $4.46$\% vs. $53.23$\% in Table \ref{results_SuperM2M_6ch} and $7.02$\% vs. $10.20$\% in Table \ref{results_SuperM2M_1ch}), indicating that close-talk mixtures are not must-have for our system.

Next, we include close-talk mixtures for training.
In system $4$a-$4$h, the number of future taps for the close-talk microphone, $J_0$, is tuned to a value shared by all the training utterances, and we enumerate $J_0$ from $1$ all the way up to $8$.
In system $5$, we instead estimate $J_0$, on the fly, for each training mixture during training, using the technique described in Section \ref{estimate_future_taps_description}.
Compared with system $4$a-$4$h, system $5$ obtains better ASR performance on the real test data, indicating the effectiveness of estimating $J_0$.

In both Table \ref{results_SuperM2M_6ch} and \ref{results_SuperM2M_1ch}, comparing system $5$ and $3$, we observe that including close-talk mixtures for training SuperM2M produces better ASR performance on the real test data of CHiME-4 (e.g., $3.97$\% vs. $4.46$\% in Table \ref{results_SuperM2M_6ch} and $6.51$\% vs. $7.02$\% in Table \ref{results_SuperM2M_1ch}).

\begin{table*}[]
\scriptsize
\centering
\sisetup{table-format=2.2,round-mode=places,round-precision=2,table-number-alignment=center,detect-weight=true,detect-inline-weight=math}
\caption{SuperM2M vs. USES \cite{Zhang2023USES} and USES2 \cite{Zhang2024USES2} on CHiME-4\\(ASR model: Whisper Large v2)}
\label{chime4_comparison_with_USES}
\setlength{\tabcolsep}{2.5pt}
\resizebox{1.7\columnwidth}{!}{ %
\begin{tabular}{
l %
c %
c %
c %
c %
S[table-format=2.1,round-precision=1] %
S[table-format=2.1,round-precision=1] %
S[table-format=1.2,round-precision=2] %
S[table-format=1.3,round-precision=3] %
S[table-format=2.2,round-precision=2] %
S[table-format=2.2,round-precision=2] %
S[table-format=2.2,round-precision=2] %
S[table-format=2.2,round-precision=2] %
}
\toprule
& & & & & \multicolumn{4}{c}{SIMU Test (CH5)} & \multicolumn{4}{c}{CH5 of CHiME-4 Test Utterances} \\

\cmidrule(lr){6-9}\cmidrule(lr){10-13}

& \multirow{2}{*}[-6pt]{\begin{tabular}[c]{@{}c@{}}Cross\\reference\end{tabular}} & & \multirow{2}{*}[-6pt]{\begin{tabular}[c]{@{}c@{}}Training\\data\end{tabular}} & & {\multirow{2}{*}[-6pt]{\begin{tabular}[c]{@{}c@{}}SISDR\\(dB)$\uparrow$ \end{tabular}}} & {{\multirow{2}{*}[-6pt]{\begin{tabular}[c]{@{}c@{}}SDR\\(dB)$\uparrow$ \end{tabular}}}} & {{\multirow{2}{*}[-6pt]{\begin{tabular}[c]{@{}c@{}}WB-\\PESQ$\uparrow$ \end{tabular}}}} & & \multicolumn{2}{c}{Val. WER (\%)$\downarrow$} & \multicolumn{2}{c}{Test WER (\%)$\downarrow$} \\

\cmidrule(lr){10-11} \cmidrule(lr){12-13}

{\multirow{1}{*}[0pt]{\rotatebox[origin=c]{0}{System}}} & & Approach & & Task & & & & {\multirow{1}{*}{STOI$\uparrow$}} & {SIMU} & {REAL} & {SIMU} & {REAL} \\
\midrule
0 & - & Mixture & {-} & {$1$-channel} & 7.506217692752905 & 7.539180193284881 & 1.27 & 0.8702431371034524 & 5.489674088415618 & 5.08652333508128 & 5.81575682382134 & 6.690450338658807 \\
\midrule
1a & - & USES \cite{Zhang2023USES} & S & {$1$-channel} & {-} & {-} & {-} & {-} & {-} & {-} & {-} & 11.0 \\
1b & - & USES \cite{Zhang2023USES} & S+extra & {$1$-channel} & {-} & {-} & {-} & {-} & {-} & {-} & {-} & 7.1 \\
[0.5ex]\hdashline\noalign{\vskip 0.5ex}
2 & 7c of Table \ref{results_SuperM2M_1ch} & SuperM2M & S+R & {$1$-channel} & 16.60562385325179 & 17.418666160478683 & 2.510620386221192 & 0.9626600245742095 & 5.3727008712487905 & 4.840466298253398 & 6.136269644334161 & 5.733933095496614 \\
\midrule
3a & - & USES \cite{Zhang2023USES} & S & {$6$-channel} & {-} & 20.6 & 3.16 & 0.983 & {-} & {-} & 4.2 & 78.1 \\
3b & - & USES \cite{Zhang2023USES} & S+extra & {$6$-channel} & {-} & 19.1 & 2.95 & 0.979 & {-} & {-} & 4.1 & 85.9 \\
3c & - & USES2 \cite{Zhang2024USES2} & S & {$6$-channel} & {-} & 18.8 & 2.94 & 0.979 & {-} & {-} & 4.6 & 12.1 \\
3d & - & USES2 \cite{Zhang2024USES2} & S+extra & {$6$-channel} & {-} & {-} & {-} & {-} & {-} & {-} & {-} & 10.3 \\
[0.5ex]\hdashline\noalign{\vskip 0.5ex}
4 & 7c of Table \ref{results_SuperM2M_6ch} & SuperM2M & S+R & {$6$-channel} & 22.849200136282228 & 23.214971050443765 & 3.215787200223316 & 0.9867863526257546 & 3.384156179412714 & 3.8078334879593405 & 3.773779983457403 & 4.043224238663978 \\
\bottomrule
\multicolumn{13}{l}{\textit{Notes}: \,(a) The ``extra'' in ``SIMU+extra'' means extra $\sim$$230$ hours of simulated training data (see \cite{Zhang2023USES, Zhang2024USES2} for details).}\\
\multicolumn{13}{l}{\quad\quad\quad(b) The ``cross reference'' entry means that the other setups are the same as the ones in the referred row.}
\end{tabular}
}
\end{table*}

\subsection{SuperM2M vs. UNSSOR and M2M}

In system $6$a-$6$d of Table \ref{results_SuperM2M_6ch} and \ref{results_SuperM2M_1ch}, we respectively report the results of UNSSOR \cite{Wang2023UNSSOR} and M2M \cite{Wang2024M2MSPL}.
UNSSOR is trained in an unsupervised way directly on far-field mixtures by using the $\mathcal{L}_{\text{MC}}$ loss in (\ref{MC_loss}) but without including $\mathcal{L}_{\text{MC},0}$ (i.e., the loss on close-talk mixture).
M2M is the same as SuperM2M but without the supervised learning branch.
We can train UNSSOR and M2M on the real data alone (denoted as \textbf{R}), or on the simulated and real data combined (i.e., S+R) considering that SuperM2M is trained on S+R.
When M2M is trained on S+R, the loss function on the simulated data does not include the loss on close-talk mixture, since CHiME-4 does not provide simulated close-talk mixtures.

Since the source estimates produced by UNSSOR and M2M could suffer from frequency permutation, we employ an existing algorithm \cite{Sawada2007, Vu2010}\footnote{See \url{https://github.com/fgnt/pb_bss/blob/master/pb_bss/permutation_alignment.py}} for frequency alignment.
The algorithm exploits inter-frequency correlation of estimated source posteriors, and we compute the source posteriors based on the DNN estimates $\hat{S}_q$ and $\hat{V}_q$ respectively via $|\hat{S}_q| / \big(|\hat{S}_q| + |\hat{V}_q|\big)$ and $|\hat{V}_q| / \big(|\hat{S}_q| + |\hat{V}_q|\big)$.
In addition, since the source estimates of UNSSOR and M2M exhibit source permutation, for each evaluation metric we compute a score for each estimate and select the better score.

Comparing system $6$a-$6$d and $5$, we can see that SuperM2M obtains much better performance than UNSSOR and M2M, suggesting the effectiveness of the proposed co-learning mechanism.
The low performance of M2M and UNSSOR is possibly because the noise in CHiME-4 typically consists of an unknown number of diffuse and directional sources.
In this case, unsupervised algorithms with two hypothesized directional sources tend to get confused about which source should be the target speech source.
This problem is naturally avoided by the supervised learning mechanism in SuperM2M, which models noise sources as a single, combined source.

\subsection{Miscellaneous Results}

In system $7$a, $7$b, and $7$c of Table \ref{results_SuperM2M_6ch} and \ref{results_SuperM2M_1ch}, we switch TF-GridNet from v1 to v2, perform iSTFT-STFT projection, and apply SNR augmentation. \ZQHLRtwo{Comparing system $7$a with $7$b, we observe that applying iSTFT-STFT projection can improve ASR performance in most cases.}
From system $5$ to $7$c, although the ASR performance on the real test data gets better in Table \ref{results_SuperM2M_1ch} while worse in Table \ref{results_SuperM2M_6ch}, in both tables better ASR performance is observed on the real validation set.
We therefore use system $7$c as our default system for the rest experiments of this paper.

\subsection{SuperM2M vs. Purely Large-Scale Supervised Training}\label{comparison_with_purely_supervised_USES}

Table \ref{chime4_comparison_with_USES} compares SuperM2M with a representative line of research (USES \cite{Zhang2023USES} and USES2 \cite{Zhang2024USES2}), which trains enhancement models in a purely-supervised way on a much larger-scale simulated data of $\sim$$245$ hours, which is $\sim$$14$ times the size of the CHiME-4 SIMU+REAL data (i.e., $245/(15.1+2.9)$).

USES and USES2 have shown that, by increasing the diversity of simulated training mixtures to cover as many conditions (that could happen in real data) as possible, better enhancement can be achieved on real data.
However, on CHiME-4, compared with using unprocessed mixtures directly for recognition, they do not obtain better ASR performance, likely because the simulated training data is not representative of the real data in CHiME-4.

In comparison, SuperM2M, although trained only on the official small-scale CHiME-4 simulated and real mixtures, obtains much better ASR performance on the real data of CHiME-4 in both single- and multi-channel cases than USES, USES2, and the unprocessed mixtures.
This comparison does not suggest that purely supervised learning on large-scale simulated data is a bad idea, as it can offer an easy and feasible way for training DNNs to model speech patterns for enhancement, and building upon this strength, SuperM2M can very likely produce even better enhancement on real data.
Rather, it sounds an alarm that purely large-scale supervised learning on simulated data has a fundamental limitation incurred by using the current simulation techniques, which usually cannot simulate mixtures in a sufficiently realistic way.
In addition, it suggests the benefits of co-training the DNN on real data via M2M.
This way, the DNN, during training, can see and learn from the signal characteristics of real-recorded data, and could hence generalize better to real-recorded data.

\begin{table*}[t]
\scriptsize
\centering
\captionsetup{justification=centering}
\sisetup{table-format=2.2,round-mode=places,round-precision=2,table-number-alignment = center,detect-weight=true,detect-inline-weight=math}
\caption{ASR results in official CHiME-4 setup\\(ASR model: WavLM features + encoder-decoder model \cite{Chang2022E2EIntegration} in ESPnet)}
\label{results_CHiME4_official}
\setlength{\tabcolsep}{2.5pt}
\resizebox{1.45\columnwidth}{!}{ %
\begin{tabular}{
l %
c %
c %
c %
c %
c %
S[table-format=2,round-precision=0] %
S[table-format=2.2,round-precision=2] %
S[table-format=2.2,round-precision=2] %
S[table-format=2.2,round-precision=2] %
S[table-format=2.2,round-precision=2] %
}
\toprule
& & & & & & {\multirow{3}{*}[-12pt]{\begin{tabular}[c]{@{}c@{}}Spk.\\reinf.\\$\gamma$ (dB)\end{tabular}}} & \multicolumn{4}{c}{Official CHiME-4 Test Utterances} \\

\cmidrule(lr){8-11}

& \multirow{2}{*}[-6pt]{\begin{tabular}[c]{@{}c@{}}Cross\\reference\end{tabular}} & & & \multirow{2}{*}[-6pt]{\begin{tabular}[c]{@{}c@{}}Joint\\training\end{tabular}} & \multirow{2}{*}[-6pt]{\begin{tabular}[c]{@{}c@{}}Input\\\#mics\end{tabular}} & & \multicolumn{2}{c}{Val. WER (\%)$\downarrow$} & \multicolumn{2}{c}{Test WER (\%)$\downarrow$} \\

\cmidrule(lr){8-9} \cmidrule(lr){10-11}

{\multirow{1}{*}[0pt]{\rotatebox[origin=c]{0}{System}}} & & {\multirow{1}{*}{Approach}} & Frontend & & & & {SIMU} & {REAL} & {SIMU} & {REAL} \\

\midrule

0a & - & Mixture \cite{Chang2022E2EIntegration} & - & - & 1 & {-} & 5.93 & 4.03 & 8.25 & 4.47 \\
[0.5ex]\hdashline\noalign{\vskip 0.5ex}
0b & - & \begin{tabular}{@{}c@{}}Mixture\\(reproduced)\end{tabular} & - & - & 1 & {-} & 5.932890855 & 4.070946568 & 8.288195741 & 4.470082675 \\

\midrule

1a & - & IRIS \cite{Chang2022E2EIntegration} & Conv-TasNet & \xmark & 1 & {-} & 5.96 & 4.37 & 13.52 & 12.11 \\
1b & - & IRIS \cite{Chang2022E2EIntegration} & Conv-TasNet & \cmark & 1 & {-} & 3.16 & 2.03 & 6.12 & 3.92 \\
[0.5ex]\hdashline\noalign{\vskip 0.5ex}
2a & 7c of Table \ref{results_SuperM2M_1ch} & SuperM2M & TFGridNetv2 & \xmark & 1 & {-} & 3.392330383 & 1.836350897 & 6.569854314 & 3.03610631 \\
2b & 7c of Table \ref{results_SuperM2M_1ch} & SuperM2M & TFGridNetv2 & \xmark & 1 & 10 & \bfseries 2.400442477 & \bfseries 1.640915962 & \bfseries 4.538662682 & \bfseries 2.400859451 \\

\midrule

3a & - & MultiIRIS \cite{Masuyama2023SE} & Neural WPD & \xmark & 2 & {-} & 2.28 & 2.06 & 2.30 & 3.63 \\
3b & - & MultiIRIS \cite{Masuyama2023SE} & Neural WPD & \cmark & 2 & {-} & 2.04 & 1.66 & 2.04 & 2.65 \\
[0.5ex]\hdashline\noalign{\vskip 0.5ex}
4a & 7c of Table \ref{results_SuperM2M_6ch} & SuperM2M & TFGridNetv2 & \xmark & 2 & {-} & 1.497050147 & 1.401231608 & 2.082555098 & 1.943108038 \\
4b & 7c of Table \ref{results_SuperM2M_6ch} & SuperM2M & TFGridNetv2 & \xmark & 2 & 10 & \bfseries 1.28318584 & \bfseries 1.331170028 & \bfseries 1.881770638 & \bfseries 1.835676584 \\

\midrule

5a & - & MultiIRIS \cite{Masuyama2023SE} & Neural WPD & \xmark & 6 & {-} & 1.19 & 1.32 & 1.29 & 1.85 \\
5b & - & MultiIRIS \cite{Masuyama2023SE} & Neural WPD & \cmark & 6 & {-} & 1.22 & 1.33 & \bfseries 1.24 & 1.77 \\
[0.5ex]\hdashline\noalign{\vskip 0.5ex}
6a & 7c of Table \ref{results_SuperM2M_6ch} & SuperM2M & TFGridNetv2 & \xmark & 6 & {-} & \bfseries 0.829646017 & 1.257420996 & 1.368135973 & 1.606800878 \\
6b & 7c of Table \ref{results_SuperM2M_6ch} & SuperM2M & TFGridNetv2 & \xmark & 6 & 10 & \bfseries 0.833333333 & \bfseries 1.231608835 & 1.372805379 & \bfseries 1.578775281 \\

\midrule

\rowcolor{shadecolor} 7 & - & \begin{tabular}{@{}c@{}}Close-talk\\Mixture\end{tabular} & - & - & - & {-} & {-} & 1.143109996 & {-} & 1.490027558 \\

\bottomrule
\multicolumn{10}{l}{\textit{Notes}: The best scores are highlighted in bold in the $1$-, $2$- and $6$-channel cases separately.}
\end{tabular}
}
\end{table*}

\subsection{Breaking Out to New Highs on CHiME-4 ASR Tasks}

Table \ref{results_CHiME4_official} reports the ASR performance of SuperM2M based on the official ASR setup of CHiME-4, using the ASR model proposed in \cite{Chang2022E2EIntegration} and still following the evaluation pipeline in Fig. \ref{enh_asr_pipeline_figure}.
Comparing system $0$b with $0$a, we observe that we have successfully reproduced the ASR system proposed in \cite{Chang2022E2EIntegration}.

SuperM2M, despite not jointly trained with the ASR model, achieves a new state-of-the-art on the REAL test set in each of the $1$-, $2$- and $6$-channel tasks, significantly outperforming the previous best obtained by IRIS \cite{Chang2022E2EIntegration} and MultiIRIS \cite{Masuyama2023SE} (e.g., in the $1$-channel case $3.04$\% vs. $3.92$\% WER in $2$a and $1$b, in the $2$-channel case $1.94$\% vs. $2.65$\% WER in $4$a and $3$b, and in the $6$-channel case $1.61$\% vs. $1.77$\% WER in $6$a and $5$b).
IRIS jointly trains a Conv-TasNet based monaural enhancement model, a WavLM based ASR feature extractor, and an encoder-decoder transformer based ASR model.
MultiIRIS, building upon IRIS, replaces the Conv-TasNet module with a DNN based weighted power minimization distortionless response (WPD) beamformer.
From system $1$a vs. $1$b, $3$a vs. $3$b, and $5$a vs. $5$b, we observe that, without joint training, IRIS and MultiIRIS often obtain clearly worse ASR performance, especially in the $1$- and $2$-channel cases.
These results further indicate the effectiveness and potential of SuperM2M on real data, as it decouples enhancement and ASR and does not employ joint training.

\subsection{Effects of Speaker Reinforcement}

In system $8$a-$8$c of Table \ref{results_SuperM2M_6ch} and \ref{results_SuperM2M_1ch}, where the Whisper ASR model is used for recognition, we apply speaker reinforcement with the energy level $\gamma$ between the enhancement output and input mixture tuned based on the set of $\{10, 15, 20\}$ dB.
Better ASR performance is observed in the $1$-channel case but not in the $6$-channel case, possibly because the enhanced speech is already reliable in the $6$-channel case, rendering speaker reinforcement not necessary.

In Table \ref{results_CHiME4_official}, where the ASR system proposed in \cite{Chang2022E2EIntegration} is used for recognition, applying speaker reinforcement in system $2$b, $4$b and $6$b respectively outperforms $2$a, $4$a and $6$a, pushing down the WER on the real test set to $2.40$\%, $1.84$\% and $1.58$\%.

\subsection{Comparison with Using Close-Talk Mixtures for ASR}

In system $7$ of Table \ref{results_CHiME4_official} and system $9$ of Table \ref{results_SuperM2M_6ch} and \ref{results_SuperM2M_1ch}, we provide the ASR results of using close-talk mixtures for decoding.
We observe that our proposed $6$-channel system obtains ASR results comparable to using close-talk mixtures for decoding.
This further indicates the effectiveness of SuperM2M and our robust ASR system.

\ZQHLRtwo{
\subsection{Whisper vs. ESPnet ASR Model under Same Text Norm.}\label{whisper_vs_ESPnet_same_text_norm}

In previous tables, the WER results of using the Whisper ASR model and using the ESPnet ASR model cannot be directly compared, due to the use of different text normalization.
As a side result, we compare the ESPnet ASR model with the Whisper Large v2 model, by traininig and evaluating the ESPnet ASR model using the same text normalizer as Whisper.
Table \ref{whisper_vs_ESPnet} reports the results.
We observe that the ESPnet ASR model obtains clearly better WERs than Whisper on the CHiME-4 dataset.
}

\begin{table*}[t]
\scriptsize
\centering
\captionsetup{justification=centering}
\sisetup{table-format=2.2,round-mode=places,round-precision=2,table-number-alignment = center,detect-weight=true,detect-inline-weight=math}
\caption{ASR results on CHiME-4 obtained by using\\Whisper Large v2 vs. ESPnet ASR model \cite{Chang2022E2EIntegration} under same Whisper text normalization.}
\label{whisper_vs_ESPnet}
\setlength{\tabcolsep}{3pt}
\resizebox{1.35\columnwidth}{!}{ %
\begin{tabular}{
l %
c %
c %
c %
c %
S[table-format=2.2,round-precision=2] %
S[table-format=2.2,round-precision=2] %
S[table-format=2.2,round-precision=2] %
S[table-format=2.2,round-precision=2] %
}
\toprule
& & & & & \multicolumn{4}{c}{Official CHiME-4 Test Utterances} \\

\cmidrule(lr){6-9}

& \multirow{2}{*}[-6pt]{\begin{tabular}[c]{@{}c@{}}Cross\\reference\end{tabular}} & & \multirow{2}{*}[-6pt]{\begin{tabular}[c]{@{}c@{}}ASR model\\(same Whisper text norm.)\end{tabular}} & \multirow{2}{*}[-6pt]{\begin{tabular}[c]{@{}c@{}}Input\\\#mics\end{tabular}} & \multicolumn{2}{c}{Val. WER (\%)$\downarrow$} & \multicolumn{2}{c}{Test WER (\%)$\downarrow$} \\

\cmidrule(lr){6-7} \cmidrule(lr){8-9}

{\multirow{1}{*}[0pt]{\rotatebox[origin=c]{0}{System}}} & & {\multirow{1}{*}{Approach}} & & & {SIMU} & {REAL} & {SIMU} & {REAL} \\

\midrule

1a & - & Mixture & Whisper Large v2 & 1 & 7.429816069699903 & 4.957444233794522 & 10.96980976013234 & 7.693500853109973 \\
1b & - & Mixture & ESPnet ASR Model \cite{Chang2022E2EIntegration} & 1 & \bfseries 6.937721845756696 & \bfseries 4.9372756242184665 & \bfseries 9.599875930521092 & \bfseries 5.4237112869034695 \\

\midrule

2a & 7c of Table \ref{results_SuperM2M_1ch} & SuperM2M & Whisper Large v2 & 1 & 7.046627944498225 & 4.783994191440442 & 11.342018196856907 & 6.023473450183548 \\
2b & 7c of Table \ref{results_SuperM2M_1ch} & SuperM2M & ESPnet ASR Model \cite{Chang2022E2EIntegration} & 1 & \bfseries 3.956921587608906 & \bfseries 2.29518776975515 & \bfseries 7.83705541770058 & \bfseries 3.650276614445996 \\

\midrule

3a & 7c of Table \ref{results_SuperM2M_6ch} & SuperM2M & Whisper Large v2 & 6 & 3.384156179412714 & 3.8078334879593405 & 3.773779983457403 & 4.043224238663978 \\
3b & 7c of Table \ref{results_SuperM2M_6ch} & SuperM2M & ESPnet ASR Model \cite{Chang2022E2EIntegration} & 6 & \bfseries 1.1898999677315263 & \bfseries 1.629623653745311 & \bfseries 1.4526468155500414 & \bfseries 1.9026937593712837 \\

\bottomrule
\end{tabular}
}
\end{table*}

\section{Conclusion}

We have proposed to adapt UNSSOR and M2M training for neural speech enhancement, where the models can be trained on real-recorded far-field mixtures in an unsupervised way, and on real-recorded close-talk and far-field mixture pairs in a weakly-supervised way.
To improve UNSSOR and M2M training, we have proposed SuperM2M, a novel co-learning algorithm that trains neural speech enhancement models by alternating between supervised training on simulated data and UNSSOR/M2M training on real data.
Evaluation results on the challenging CHiME-4 benchmark show the effectiveness of SuperM2M for speech enhancement and robust ASR.
Future research will modify and evaluate SuperM2M on conversational speech separation and recognition.

Our study, we think, provides illuminating findings towards improving the generalizability of modern neural speech enhancement models to real-recorded data, since it, for the first time since the introduction of the challenging and representitive CHiME-4 benchmark nearly a decade ago, shows that, on the real mixtures of CHiME-4, feeding in the immediate outputs of neural speech enhancement models for ASR decoding can produce remarkable improvement over feeding in unprocessed mixtures and neural beamforming results, breaking out to new highs in ASR performance even though joint frontend-backend training is not employed and even if the ASR backend, which leverages strong self-supervised learning representations, is a very strong one.
This success is realized by SuperM2M, which trains enhancement models not only on simulated data but also on real data, and through our accumulative efforts on complex spectral mapping \cite{Wang2020chime, Wang2020css, Tan2022NSF}, loss functions dealing with implicit magnitude-phase compensation \cite{Wang2021compensation}, FCP \cite{Wang2021FCPjournal, Wang2021FCPwaspaa}, TF-GridNet \cite{Wang2022GridNetjournal}, UNSSOR \cite{Wang2023UNSSOR}, USDnet \cite{Wang2024USDnet}, and M2M \cite{Wang2024M2MSPL}, which have firmly built up the foundation of SuperM2M.

We point out that nearly all the current supervised neural speech enhancement algorithms can be seamlessly integrated with SuperM2M to improve their generalization abilities, by including real-recorded close-talk and far-field mixture pairs, or far-field mixtures alone if close-talk mixtures are unavailable, for M2M training.
This indicates that SuperM2M can ride on the development of large-scale supervised neural speech enhancement models trained on simulated data, and has strong potential to grow into a representative algorithm for training speech enhancement models directly on real-recorded data.

In closing, we would like to highlight the learning-based methodology for solving blind deconvolution problems, which has been developed along our line of research on FCP \cite{Wang2021FCPjournal, Wang2021FCPwaspaa}, UNSSOR \cite{Wang2023UNSSOR}, M2M \cite{Wang2024M2MSPL}, and SuperM2M.
By training DNNs in an un-, weakly- or semi-supervised way to estimate sources, filter estimation becomes differentiable so that the DNNs can be trained to optimize mixture-constraint losses to realize separation.
Based on the challenging real-recorded mixtures in CHiME-4 and through the integration with supervised learning, this paper, for the first time, has demonstrated that this methodology is effective for real-recorded data, and is also effective at neural speech enhancement.
Considering that blind deconvolution broadly exists in many application domains, we expect the methodology to be also effective in similar applications and generate broader impact beyond speech enhancement.

\section{Acknowledgments}

We would like to thank Dr. Xuankai Chang at Carnegie Mellon University for constructive discussions on robust ASR.

\bibliographystyle{elsarticle-num.bst}
\bibliography{references.bib}

\end{document}